\documentclass[11pt,a4paper]{article}
\usepackage{amsmath}
\usepackage[dvipdfmx,hiresbb]{graphicx}
\usepackage{xcolor}
\usepackage{mathrsfs,mathtools}
\usepackage{physics,amssymb}
\usepackage{siunitx}
\usepackage{bm}
\usepackage{braket}
\usepackage{listings}

\usepackage{cases}
\usepackage{comment}
\usepackage{soul}
\usepackage{cancel}
\usepackage{cases}
\usepackage[utf8]{inputenc}
\usepackage{url}
\usepackage{float}
\usepackage{longtable}
\usepackage[normalem]{ulem}
\usepackage{xspace}
\usepackage{aas_macros}
\usepackage{autobreak}

\definecolor{Green}{HTML}{026E33}

\usepackage{ulem}
\DeclareRobustCommand{\Erase}{\bgroup\markoverwith{\textcolor{red}{\rule[.5ex]{2pt}{0.4pt}}}\ULon}

\usepackage{subcaption}
\usepackage{empheq}
\usepackage{here}
\usepackage{pdfpages}
\usepackage{pgfplots}

\setstcolor{red}
\usepackage{jcappub}
\hypersetup{colorlinks=true
,urlcolor=DARKBLUE
,anchorcolor=DARKBLUE
,citecolor=DARKBLUE
,filecolor=DARKBLUE
,linkcolor=DARKBLUE
,menucolor=DARKBLUE
,linktocpage=true
,pdfproducer=medialab
}

\definecolor{MONZA}{HTML}{CF000F}
\definecolor{DARKBLUE}{HTML}{00008b}
 \definecolor{DARKBLUE}{rgb}{0,0,0.7} 
\definecolor{DARKMAGENTA}{HTML}{8b008b}
\definecolor{DARKCYAN}{HTML}{008B8B}
\definecolor{DARKORANGE}{HTML}{FF8C00}

\definecolor{MAGENTA}{HTML}{FF00FF}

\begin{document}

\title{Numerical simulation of type II primordial black hole formation}

\author[a]{Koichiro Uehara,}
\author[a]{Albert Escriv\`{a},}
\author[b]{Tomohiro Harada,}
\author[a]{~~~~~~~~~~~Daiki Saito,}
\author[a]{and Chul-Moon Yoo}

\affiliation[a]{Division of Science, Graduate School of Science, Nagoya University, Nagoya 464-8602, Japan}
\affiliation[b]{Department of Physics, Rikkyo University, Toshima, Tokyo 171-8501, Japan}

\emailAdd{uehara.koichiro.p8@s.mail.nagoya-u.ac.jp}
\emailAdd{escriva.manas.albert.y0@a.mail.nagoya-u.ac.jp}
\emailAdd{harada@rikkyo.ac.jp}
\emailAdd{saito.daiki.g3@s.mail.nagoya-u.ac.jp}
\emailAdd{yoo.chulmoon.k6@f.mail.nagoya-u.ac.jp}

\date{\today}
\abstract{
This study investigates the formation of  primordial black holes (PBHs) resulting from extremely large amplitudes of initial fluctuations in a radiation-dominated universe. 
We find that, for a sufficiently large initial amplitude, the configuration of trapping horizons shows characteristic structure due to the existence of bifurcating trapping horizons.
We call this type of configuration of the trapping horizons type B PBH, while the structure without a bifurcating trapping horizon type A PBH. 
As shown in Ref.~\cite{PhysRevD.83.124025}, in the matter-dominated universe, the type B PBH can be realized by the type II initial fluctuation, which is characterized by a non-monotonic areal radius as a function of the radial coordinate (throat structure) in contrast with the standard case, type A PBH with a monotonic areal radius (type I fluctuation).
Our research reveals that a type II fluctuation does not necessarily result in a type B PBH in the radiation-dominated case. 
We also find that for an initial amplitude well above the threshold value, the resulting PBH mass may either increase or decrease with increasing the initial amplitude, depending on its specific profile rather than its fluctuation type.
}

\arxivnumber{2401.06329}

\maketitle
\flushbottom
\section{Introduction}
Primordial black holes (PBHs) play a crucial role in exploring the early universe, particularly in investigating inhomogeneities whose scales are much smaller than those of the fluctuations probed by the cosmological microwave background. 
Since PBHs are remnants of nonlinear inhomogeneity in the primordial universe following the inflationary era, they are unique probes of statistical properties of the small-scale primordial inhomogeneity. 
They are considered as one of the candidates for dark matter and can be detected as black hole (BH) binaries by gravitational wave interferometers. 
According to the standard scenario, PBHs were formed through the gravitational collapse of significantly high-density regions during the radiation-dominated era, which had been generated in the inflationary period \cite{Zeldovich:1967lct,10.1093/mnras/152.1.75,10.1093/mnras/168.2.399,1978SvA....22..129N,1980AZh....57..250N}.

The mechanism of PBH formation can be briefly understood by considering a model of the gravitational collapse in the flat Friedmann--Lemaître--Robertson--Walker (FLRW) universe (see, e.g., Refs.~\cite{PhysRevD.88.084051,galaxies10060112}). 
A criterion for the collapse has been first proposed by Carr \cite{osti_4109902} as $\delta \gtrsim w = 1/3$, where $\delta$ is the amplitude of the density perturbation, and $w$ is the coefficient of the assumed linear equation of state $p=w\rho$ with $p$ and $\rho$ being the pressure and the energy density, respectively. A more accurate threshold value has been explored by numerically analyzing general relativistic nonlinear dynamics \cite{Niemeyer:1999ak, PhysRevD.60.084002, Musco_2005,PhysRevD.88.084051, Nakama_2014,Nakama_2014double,osti_4109902, Polnarev_2007, Musco:2008hv,Musco:2012au,ESCRIVA2020100466,Escriva:2019phb,escriva_2021analy,Harada:2015yda,Musco:2018rwt,Musco:2020jjb}.

In cosmological perturbation theory, the growth of the perturbation is often described by the evolution of the curvature perturbation $\zeta$, which represents the growing adiabatic modes. 
To perform numerical simulations of PBH formation, one must establish the initial conditions for the spacetime geometry and the fluid in the system. 
This involves setting the initial profile of curvature fluctuations at the super-horizon scale before the horizon entry to be consistent with the assumption of the growing adiabatic modes. 
These analytical constructions are typically achieved using the gradient expansion techniques \cite{PhysRevD.60.084002, Polnarev_2007,Harada:2015yda}.

In a spherically symmetric system, we introduce a type II primordial fluctuation characterized by a non-monotonic behavior of the areal radius $R=\sqrt{A/4\pi}$, where $A$ is the area of the 2-sphere as a function of the radial coordinate $r$ with constant $t$.
These fluctuations exhibit distinct neck-like structures, with a minimal areal radius at $\partial_r R = 0$.
In cosmological perturbation theory, the areal radius $R$ is often expressed via the curvature fluctuation $\zeta$, $R\propto r e^\zeta$.
The type II fluctuation is only realized by a sufficiently large amplitude of $\zeta$ with the scale fixed.

Kopp, Hofmann, and Weller (KHW)~\cite{PhysRevD.83.124025} investigated PBH formation associated with the type II fluctuations by employing the Lema\^{i}tre--Tolman--Bondi (LTB) solution, which is the analytic solution for the dust fluid ($w=0$). 
They investigated the causal structure of the spacetime generated from a specific profile of the type II fluctuation. They explicitly showed that it describes a spacetime of BH formation, which they call type II PBH. 
In addition, they revealed distinct properties of the type II PBH from the type I PBH, which arises from typical type I fluctuations with the amplitude above the threshold. 
Carr and Harada~\cite{PhysRevD.91.084048} analytically discussed type II PBHs with more general equations of state using simplified models. Type I/II fluctuations are also important because the threshold of these two cases gives a possible maximum value of the averaged density perturbation at horizon entry in the simplified models~\cite{PhysRevD.83.124025,PhysRevD.91.084048}.
We aim to investigate the dynamics and the spacetime structure of type II PBH formation in the radiation-dominated universe, namely, to clarify the impact of the existence of the radiation pressure. 

When we consider the abundance of PBHs, the contribution from type II fluctuations is often neglected because they are statistically unlikely to be generated, and the abundance of type II PBHs originating from them is typically suppressed. 
However, as mentioned in~\cite{Gow:2022jfb}, they can be significant when the statistics of primordial fluctuations are highly non-Gaussian. 
Moreover, in the context of a single-field inflationary model, it has been recently discovered in Ref.~\cite{escriva2023formation} that type II fluctuations surround the bubbles formed when the inflaton overshoots the barrier. 
As non-Gaussianities increase, the production of vacuum bubbles becomes dominant, leading to a mass function and abundance primarily driven by these fluctuations. 
In addition, a similar BH formation mechanism due to the birth of a baby universe with a tunneling process during inflation has been reported in Refs.~\cite{Deng_2017, Deng_2017b}. 
The baby universe is connected to our universe through a throat inside the BH, like the neck structure in the type II fluctuation.

For example, in the Misner--Sharp formalism~\cite{PhysRev.136.B571} (see also Refs.~\cite{PhysRevD.83.124025, PhysRevD.74.084013}), the time evolution of the type II fluctuations via a usual areal radial coordinate system such as in Ref.~\cite{Harada:2015yda}, the neck throat structure would be complicated. 
The evolution equations suffer from the trouble associated with $0/0$ term at the neck radius satisfying $\partial_rR=0$ when we consider type II fluctuations. 
To overcome this problem, it will be necessary to conveniently modify the equations, as done in Ref.~\cite{Deng_2017} for another scenario of PBH formation from the collapse of vacuum bubbles.

In this paper, we use the simulation code developed from a 3+1 dimensional simulation code \cite{Yoo_2019} with BSSN \cite{shibata1995evolution, PhysRevD.59.024007} formalism with CARTOON method \cite{doi:10.1142/S0218271801000834} so that we can avoid the same problem.

This paper is organized as follows.
Section \ref{sec:Initialdata} describes the setup and numerical code of spherically symmetric numerical simulation of PBH formation. 
In Sec.~\ref{subsec:evol_density}, we present the simulation results, highlighting the time evolution of fluctuations.
Then, we explore the formation of trapping horizons, including the BH horizon in Sec.~\ref{subsec:AHs}, and give an interpretation of the resulting trapping horizon configurations in Sec.~\ref{subsec:3zonePic}.
In Sec.~\ref{subsec:PBHmass}, we examine the evolution of PBH mass.
In Sec.~\ref{sec:profile_dep}, to check the profile dependence of the PBH mass, we explore PBH formation from another initial profile that models the overdense region, as described by KHW.
Finally, we offer our conclusions in the last section.
Appendix A provides the conformal diagrams of the exact three-zone solutions for PBH formation. Appendix B provides the derivation of the curvature perturbation for the three-zone model. 
Appendix C proves the appearance of two peaks in the compaction function and the averaged density perturbation for the type II fluctuation.

Throughout the paper, we use the geometrical unit in which the speed of light and the gravitational Newton constant are unity, that is, $c = G =1$.

\section{Initial data setting}\label{sec:Initialdata}
We start our simulation from the time when the length scale of the fluctuation is much larger than the Hubble length $H_\text{b}^{-1}$, that is, $k^{-1} \gg (a H_\text{b})^{-1}$ with $k$, $a=a(t)$ and $H_\text{b}:=(da/dt)/a$ being the characteristic comoving wave number, the scale factor and the Hubble parameter of the background flat FLRW universe, respectively. 
We obtain the initial condition by applying the cosmological long-wavelength approximation, in which the small parameter $\epsilon:=k/(a H_{\rm b})$ is introduced\footnote{
    One would use the other expression, $\epsilon:=1/(H_{\rm b} R_{\rm m})$ with $R_{\rm m}\propto a/k$ being the areal radius, where $\mathcal{C}_{\rm SS}(r)$ takes a maximum.
}, and the zeroth order spatial metric is given by \cite{Lyth_2005}
\begin{equation}
    \gamma_{ij} = a^2(t) e^{2\zeta(x^k)} f_{ij}, \,
\end{equation}
with $f_{ij}$ being the reference 3-metric of the background flat geometry, and $\partial_i \zeta /(aH_{\rm b})=\mathcal O(\epsilon)$. 
We note that the function $\zeta$ is a time-independent arbitrary function of the spatial coordinates $x^k$. 
Once we initially set $\zeta$ as a function of the spatial coordinates $x^i$ and fix the gauge conditions, we can successively obtain the growing mode solution up to the following leading terms in this expansion \cite{PhysRevD.60.084002, Harada:2015yda}. 
These long-wavelength solutions are used as the initial data for the time evolution. 
In this paper, we initially fix the gauge by imposing the constant mean curvature slice (uniform Hubble slice) and the normal threading, for which the trace of the extrinsic curvature of the time slice is uniformly given by $-3H_{\rm b}$ and the shift vector is equal to zero.

In a spherically symmetric system, the line element at the zeroth order in powers of $\epsilon$ is denoted as
\begin{equation}
    ds^2 = -dt^2 + a^2(t) e^{2\zeta(r)} \qty(dr^2 + r^2 d \Omega^2),
\end{equation}
where $d\Omega^2$ represents the line element of a 2-sphere.
We can see that the areal radius is given by $R = a r e^\zeta $ in this line element.
Throughout this paper, to resolve the fine structure, such as the trapping horizon formation near the center $r=0$, we use the scale-up coordinate $z$, which is defined as
\begin{equation}
    r=z-\frac{\eta}{1+\eta}\frac{L}{\pi}\sin\left(\frac{\pi}{L}z\right)
    \label{eq:scale-up_coordinate}
\end{equation}
with a parameter $\eta=20$~\cite{Yoo_2019}. 
In this coordinate, the resolution in the central region is $1+2\eta$ times finer than that near the boundary $z=r=L$;
\begin{align}
    \left.\dv{r}{z}\right|_{r=L} = \qty(1+2\eta)\left.\dv{r}{z}\right|_{r=0}.
\end{align}

Up through Sec.~\ref{subsec:PBHmass}, we use the following Gaussian-shaped profile in the numerical simulation~\cite{Yoo_2019},
\begin{equation}\label{eq:Gprofile}
    \zeta(r) = \mu e^{-\frac{1}{2}k^2 r^2}W(r), 
\end{equation}
where $\mu$ is the amplitude of the fluctuation and $W(r)$ is a window function~\cite{Yoo_2019} introduced to eliminate the tail of the Gaussian function and impose $\zeta=0$ at the outer boundary of the numerical domain.\footnote{
    The explicit functional form~\cite{Yoo_2019} is 
    \begin{empheq}[left={W(r)=\empheqlbrace}]{align}
        1  \quad &\text{for}\quad 0 \leq r \leq r_\text{w}, \notag\\
        1 - \frac{\qty((r_\text{w}-L)^6 - (L-r)^6)^6}{(r_\text{w} - L)^{36}} \quad&\text{for}\quad r_\text{w} \leq r \leq L, \\
        0 \quad&\text{for}\quad L \leq r  \notag,
    \end{empheq}
    where $r_{\rm w}$ is set to $0.8L$.
}

\begin{figure}[h]
    \centering
    \hspace{-60pt}
    \begin{minipage}[b]{0.45\linewidth}
        \centering
        \includegraphics[width=1.2\linewidth]{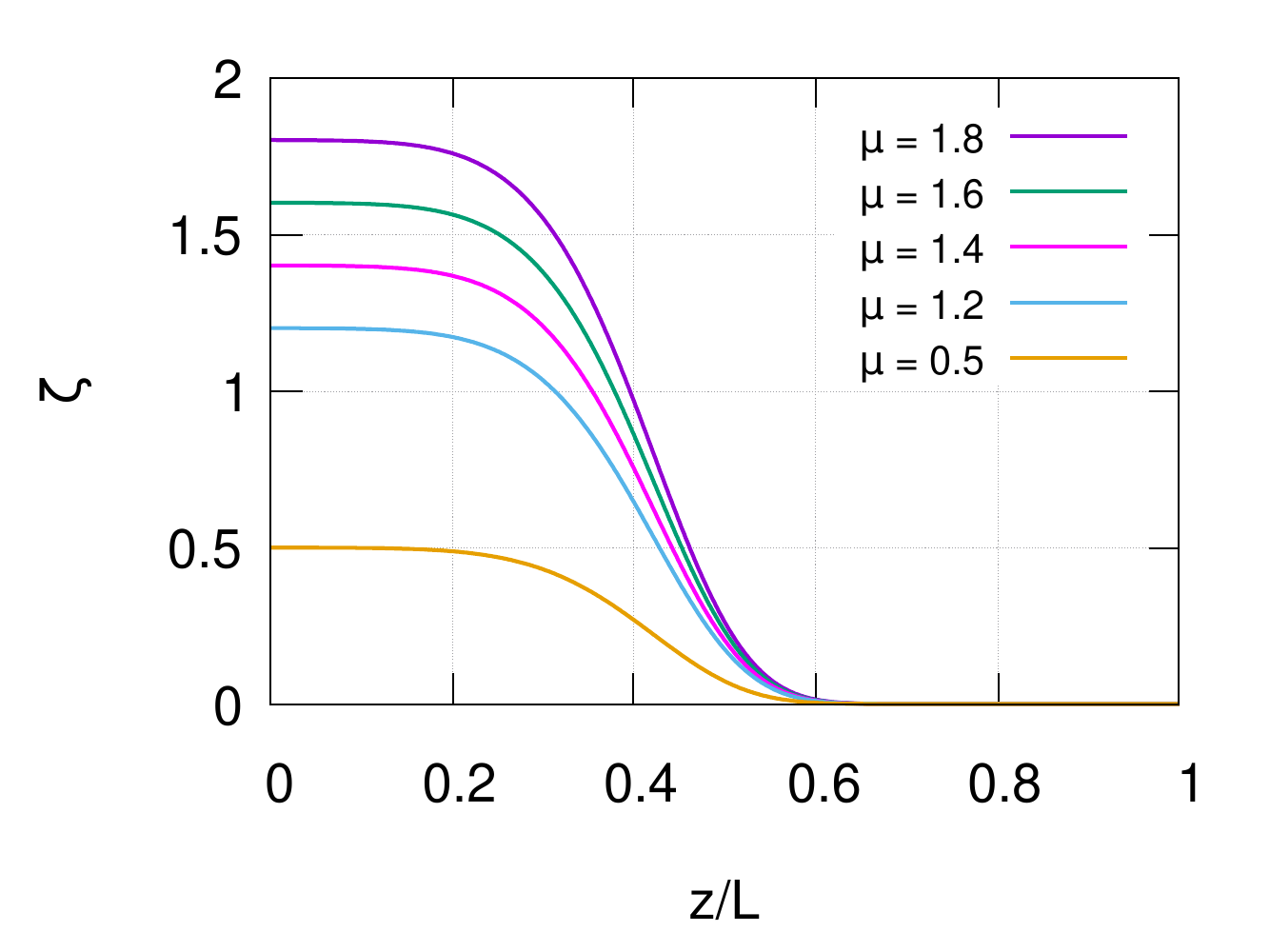}
    \end{minipage}
    \hspace{20pt}
    \begin{minipage}[b]{0.45\linewidth}
        \centering
        \includegraphics[width=1.2\linewidth]{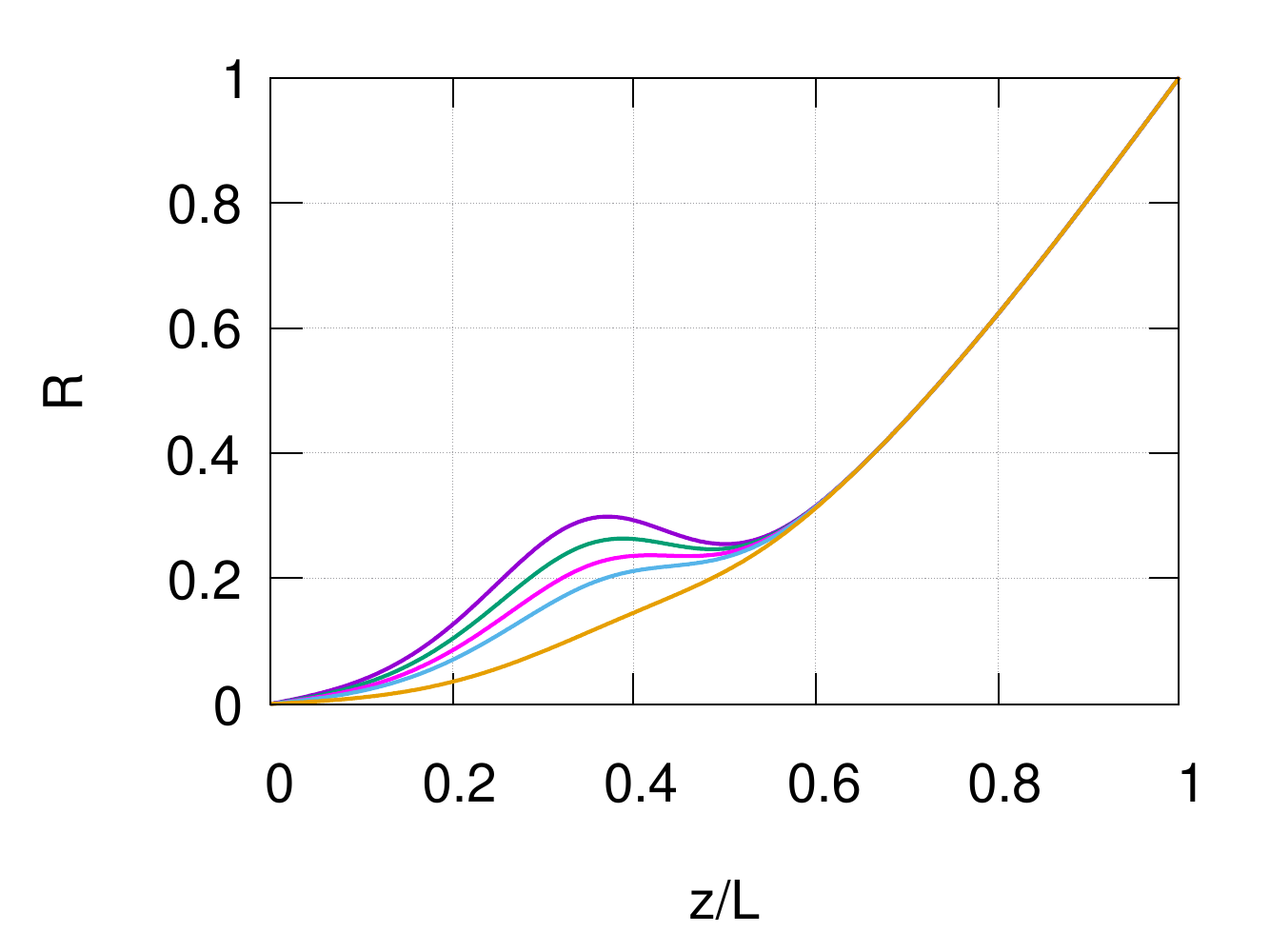}
    \end{minipage}
    \caption{
        In the left panel, the vertical and horizontal axes are the functional forms for the curvature perturbation $\zeta(r)$ given by Eq.~\eqref{eq:Gprofile} and the scale-up coordinate $z/L$ defined in Eq.~\eqref{eq:scale-up_coordinate}, respectively. Note that $z$ is a monotonically increasing function of $r$ and that $L$ is chosen for the computation region's length. We plot $\zeta(r)$ for the set of the amplitude parameter values $\mu=0.5, 1.2, 1.4, 1.6$ and $1.8$. In the right panel, the vertical and horizontal axes are the areal radius $R(r)$ and the scale-up coordinate $z/L$, respectively, for the initial data sets of the long-wavelength solutions generated by $\zeta(r)$ plotted in the left panel. We can see that $R(r)$ is a monotonically increasing function for $\mu=0.5$ and $1.2$, while it increases, decreases, and increases again as $r$ increases for $\mu=1.4, 1.6$, and $1.8$. In the latter cases, $R(r)$ has a maximum and a minimum with $\partial_{r}R=0$, where the minimum corresponds to the neck.
    }
    \label{fig:InitialGaussianProfile}
\end{figure}

Figure \ref{fig:InitialGaussianProfile} shows the functional form of this profile (left) and the areal radius $R$ (right). 
By taking a sufficiently large amplitude of the fluctuation $\mu$, the areal radius $R(r)$ in the overdense region becomes non-monotonic along the radial coordinate $r$, i.e., a pair of the local minimum and maximum points appears with $\partial_r R=0$ for $\mu\gtrsim 1.4$ as in Fig.~\ref{fig:InitialGaussianProfile}.
We call this structure the neck or throat structure, and KHW defined such fluctuations as type II and distinguished them from type I fluctuations, which have a monotonic areal radius \cite{PhysRevD.83.124025}.

\begin{figure}[h]
    \centering
    \hspace{-60pt}
    \begin{minipage}[b]{0.45\linewidth}
        \centering
        \includegraphics[width=1.2\linewidth]{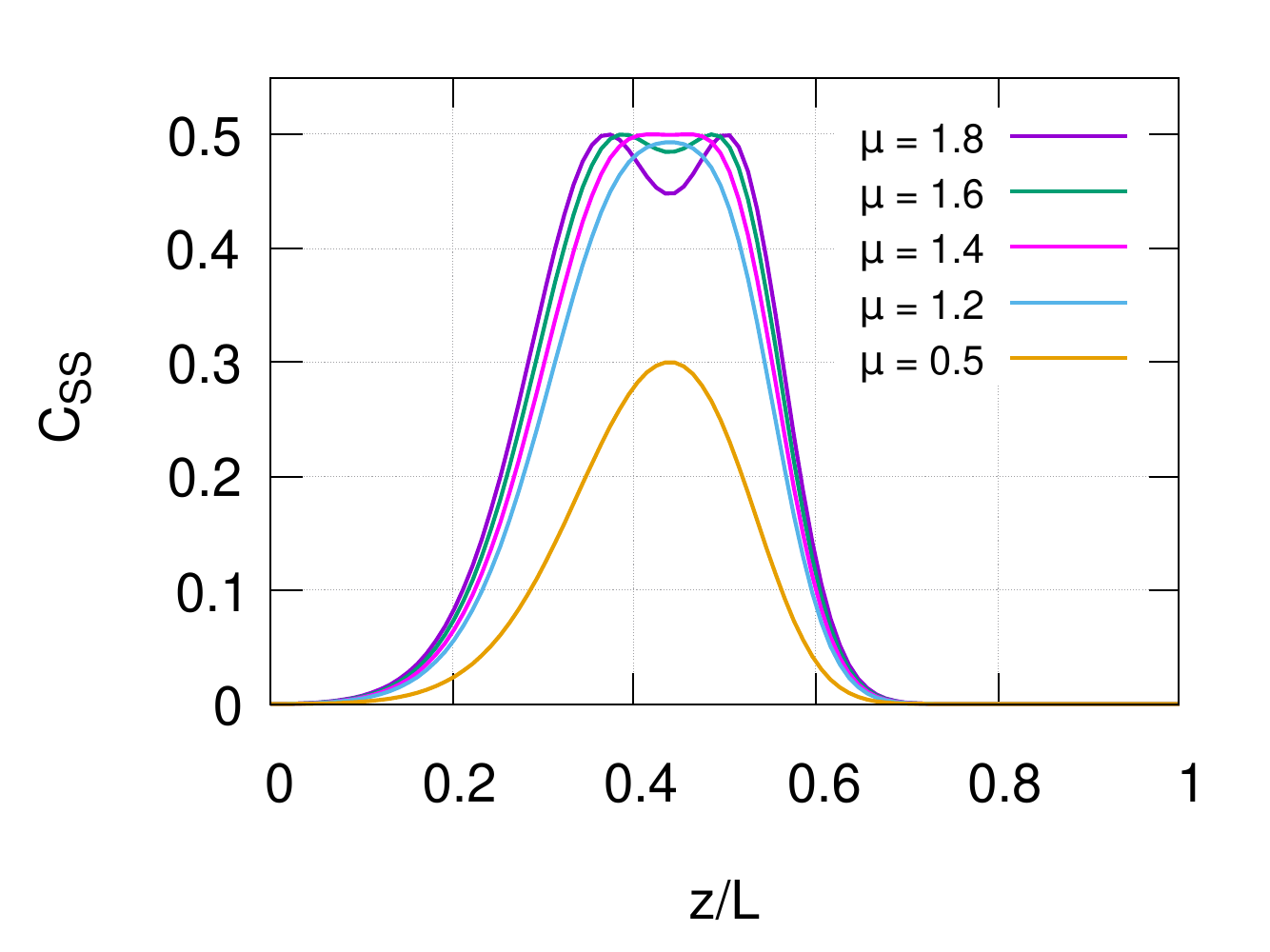}
    \end{minipage}
    \hspace{20pt}
    \begin{minipage}[b]{0.45\linewidth}
        \centering
        \includegraphics[width=1.2\linewidth]{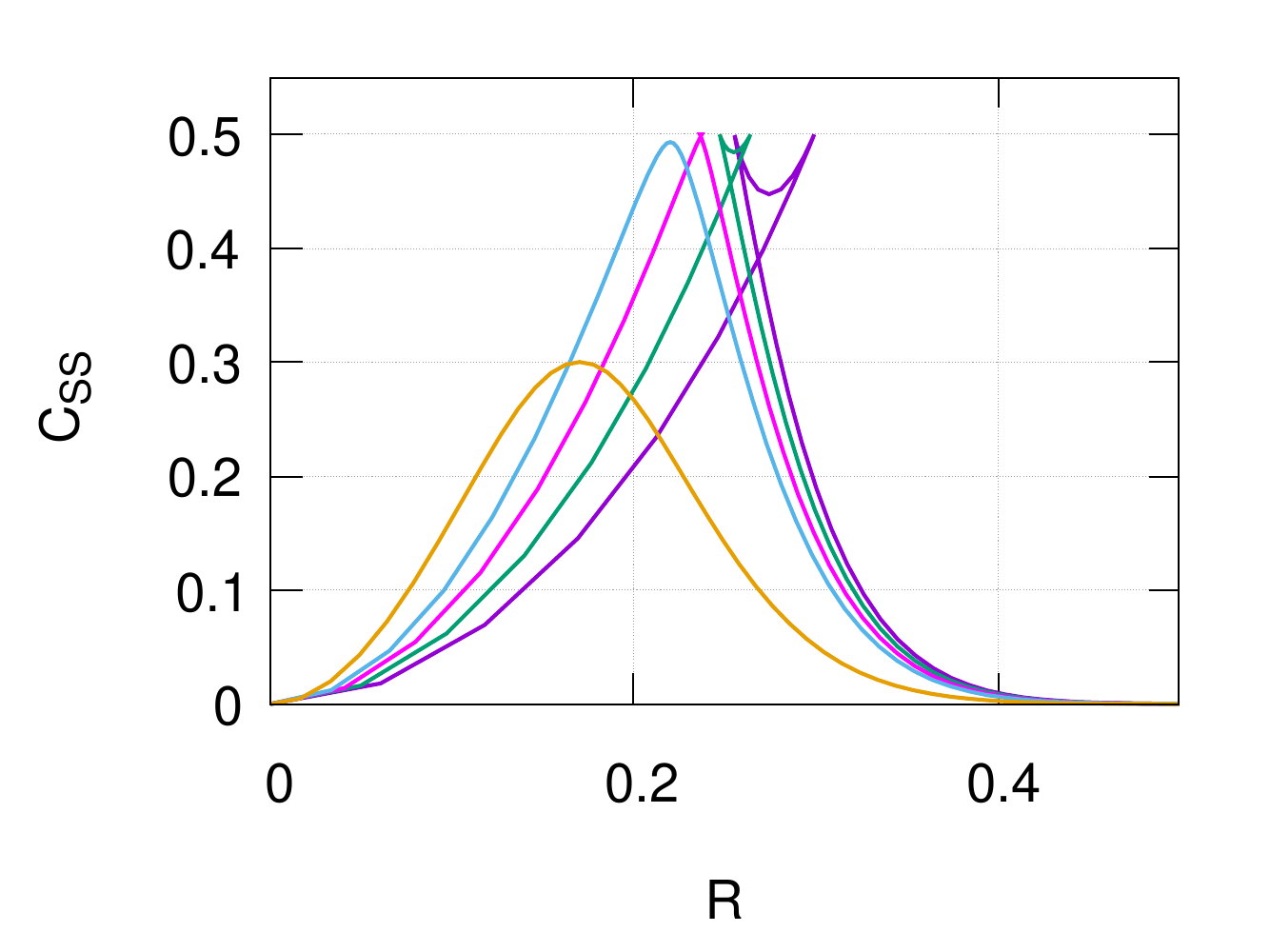}
    \end{minipage}
    \caption{
In the left panel, the vertical and horizontal axes are the compaction function $\mathcal{C}_\text{SS}(r)$ and the scale-up coordinate $z/L$, respectively, for the initial data sets of the long-wavelength solutions generated by Eq.~\eqref{eq:Gprofile} for the same set of the amplitude parameter $\mu$. We can see that $\mathcal{C}_\text{SS}(r)$ takes only one maximum, which is smaller than $1/2$, for $\mu=0.5$ and $1.2$, while it takes two distinct maxima, which are equal to $1/2$, and a single minimum between the two maxima for $\tilde{\mu}=1.4, 1.6$ and $1.8$. The latter cases correspond to type II fluctuations as shown in Appendix~\ref{sec:twoPeaks}. 
    In the right panel, the vertical and horizontal axes are the compaction function $\mathcal{C}_\text{SS}(r)$ and the areal radius $R(r)$, respectively, for the same initial data sets. We can see that $\mathcal{C}_\text{SS}(r)$ takes only the maximum or the two maxima at the larger areal radius $R$ for the larger values of $\mu$. This implies that the mass of the formed black hole increases as $\mu$ increases, as seen in Fig.~\ref{fig:mass_amp}.
    }
    \label{fig:compfunc}
\end{figure}

The feature of type II fluctuations can also be understood as two peaks\footnote{
    It is shown in Ref.~\cite{escriva2023formation} that the appearance of fluctuations of type II naturally appears via a specific single-field inflation model with a bump, but in this case, one of the peaks of $\mathcal{C}_{\rm SS}$ is hidden inside the bubble. 
}of the Shibata--Sasaki compaction function  $\mathcal{C}_\text{SS}$ at super horizon scales, which is calculated as \cite{Harada:2015yda}
\begin{equation}
    \mathcal{C}_\text{SS}(r) = \frac{1}{2}\left[1-\qty(1+ r \frac{d\zeta}{dr} )^2\right]=\frac{1}{2}\left[1-\left(\frac{r}{R} \dv{R}{r} \right)^2 \right]. 
\end{equation}
The Shibata-Sasaki compaction function was first introduced in Ref.~\cite{PhysRevD.60.084002} as an estimator for the criterion of PBH formation.
See Refs.~\cite{Harada:2023ffo,Harada:2024trx} and references therein for the initial confusion in its introduction, its usefulness, 
its relation to legitimate compaction functions in different gauges, and its geometrical origin as a conformal compactness function.
The outer peak of $\mathcal C_\text{SS}$ corresponds to the neck in the type II fluctuation, the radius at which the areal radius takes the minimum value considered relevant for PBH formation. 
We will give a general proof for two peaks and the minimum between them for the compaction function in the type II fluctuation in the Appendix~\ref{sec:twoPeaks}. 
We focus on the two peak configurations associated with type II fluctuations (see also Ref.~\cite{Escriva:2023qnq} for two peak configurations with fluctuations of type I
). 
The form of the compaction function for the specific function of $\zeta$ \eqref{eq:Gprofile} is shown in Fig.~\ref{fig:compfunc}. 
By using the position of the peak of $\mathcal{C}_\text{SS}(r)$, $r_\text{m}=\sqrt{2}/k$, we analytically obtain the threshold value $\mu_{\rm c,II}=e /2 = 1.359\cdots$ for fluctuations of type II.\footnote{
    Here, we ignore the window function $W(r)$ to estimate the location $r_\text{m}=\sqrt{2}/k$ of the peak of compaction function $\mathcal{C}_{\rm SS}(r)$ since this acts on $\zeta(r)$ only near the outer boundary, $r=r_{\rm w}=0.8L$ .
}
For $\mu > \mu_{\rm c,II}$ the compaction function takes the maximum value of $0.5$ at least one location with the radial coordinate, and the number of peak positions becomes two, $r_{m1}$ and $r_{m2}$ with $r_{m1} < r_{m2}$.
This implies that type II fluctuations always reach the collapse threshold and form PBHs irrespectively of the initial $\zeta(r)$ profile. 
The isotropic coordinate $r$ does not always give the geometrical size of the enclosed region, which can often lead to misunderstanding. 
Therefore, we also explicitly show the compaction function $\mathcal{C}_{\rm SS}$ as a function of the areal radius $R$ in the right panel of Fig.~\ref{fig:compfunc}.

To follow the time evolution of gravitational collapse from the initial profile of fluctuation, we use the numerical code COSMOS-S \cite{PhysRevLett.111.161102, PhysRevD.89.104032, PhysRevD.105.103538} based on a 3+1 dimensional code of fourth order Runge--Kutta method with BSSN formalism \cite{shibata1995evolution, PhysRevD.59.024007}.
This code is specialized for spherically symmetric systems using the CARTOON method \cite{doi:10.1142/S0218271801000834}. 
We consider the time evolution of the initial fluctuation given by the functional form of $\zeta$ \eqref{eq:Gprofile} with the equation of state parameter $w=1/3$, namely, radiation fluid. 
Here, the asymptotically spatially flat FLRW boundary condition is implemented as in Ref.~\cite{PhysRevD.60.084002}. 
However, we could not avoid violating the constraint at the outer boundary, which propagates into the bulk region. 
Therefore, we terminate the time evolution before the violation reaches the central region relevant to the PBH formation in the simulation.

We also find the constraint violation near the center for cases where an apparent horizon forms. 
Nevertheless, since we find both independent future-directed null directions inward, the constraint violation does not propagate outward. 
In practice, we excise several grids inside the apparent horizon to avoid the numerical calculation from crashing if needed.

We take the coordinate size $L$ of the numerical region ($0\leq z\leq L$ or equivalently $0\leq r\leq L$) as the unit length scale and set the initial value of the scale factor to unity. 
We fix the ratio between the initial Hubble background length $H_{\rm b}^{-1}$ and the typical scale $k^{-1}$ of the initial fluctuation as $k^{-1} = 10 H_\text{b}^{-1}$ that ensures the long-wavelength approximation. 
To avoid propagating the constraint violation, we need to take a larger region than the initial Hubble length for a longer duration of the time evolution. 
In the following, we take $H_\text{b}^{-1} = 0.01L$ or $H_\text{b}^{-1} = 0.005L$ depending on the duration of the required time evolution. 
The horizon entry time $t_H$ is defined as $k^{-1} = (a H_\text{b})^{-1}|_{t=t_H}$ throughout this paper.

\section{Spacetime structure and definition of type A/B PBHs}
\subsection{Time evolution of type I/II fluctuations}\label{subsec:evol_density}
\begin{figure}[h]
    \centering
    \hspace{-40pt}
    \begin{minipage}[b]{0.4\linewidth}
        \centering
        \includegraphics[width=1.\linewidth]{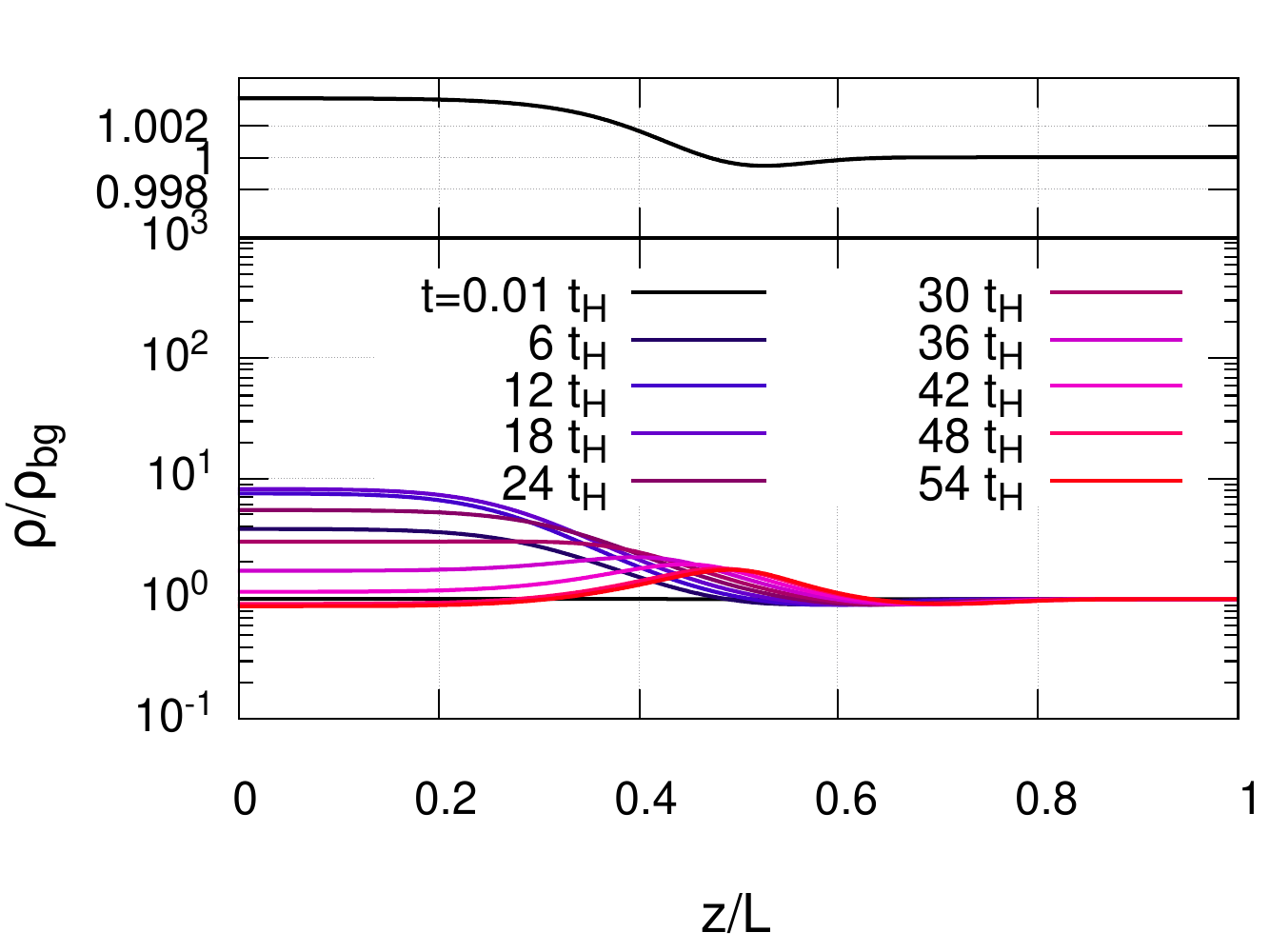} 
        \includegraphics[width=1.\linewidth]{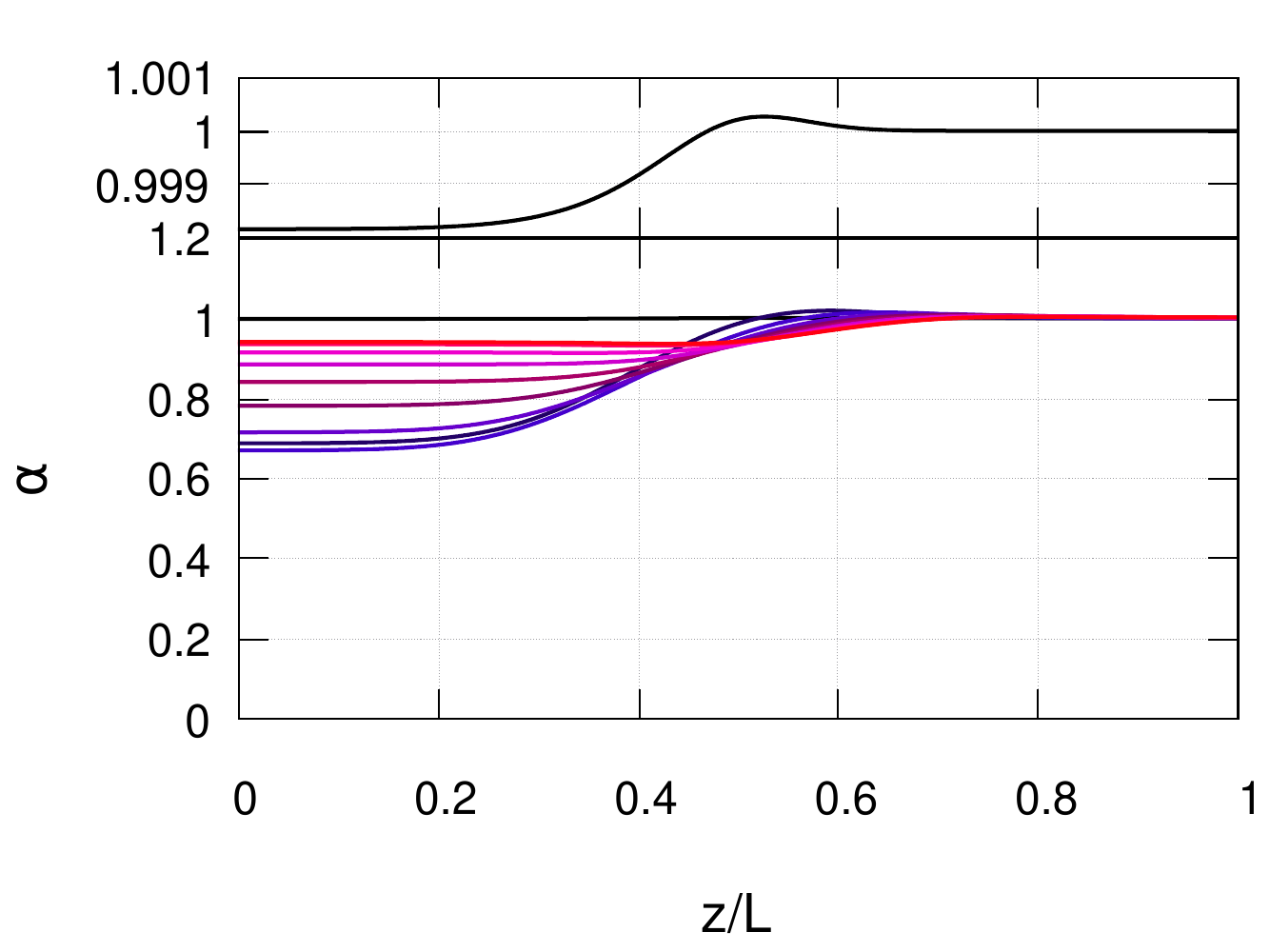}
        \subcaption{$\mu=0.5$}
    \end{minipage}
    \hspace{-40pt}
    \begin{minipage}[b]{0.4\linewidth}
        \centering
        \includegraphics[width=1.\linewidth]{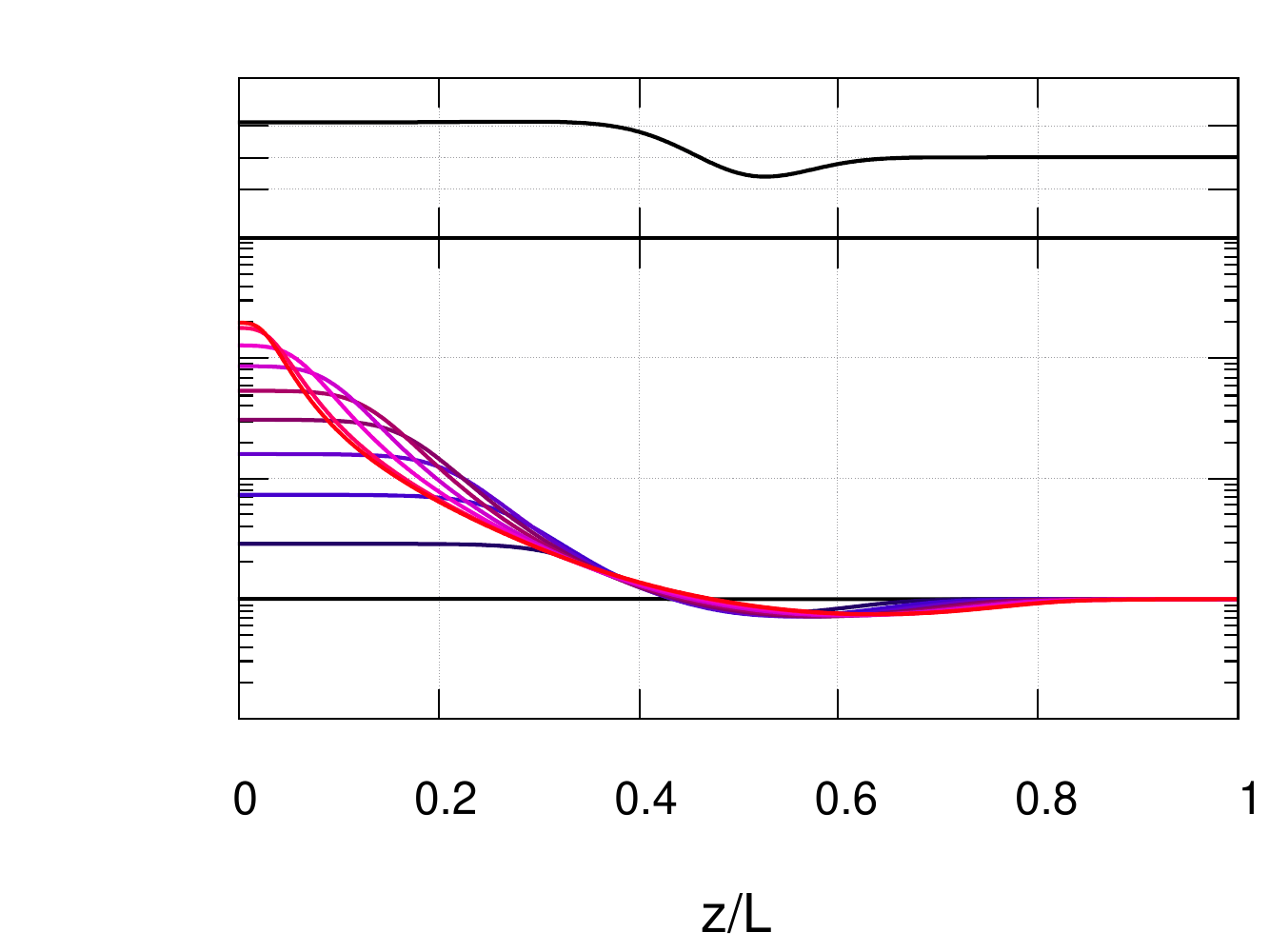}
        \includegraphics[width=1.\linewidth]{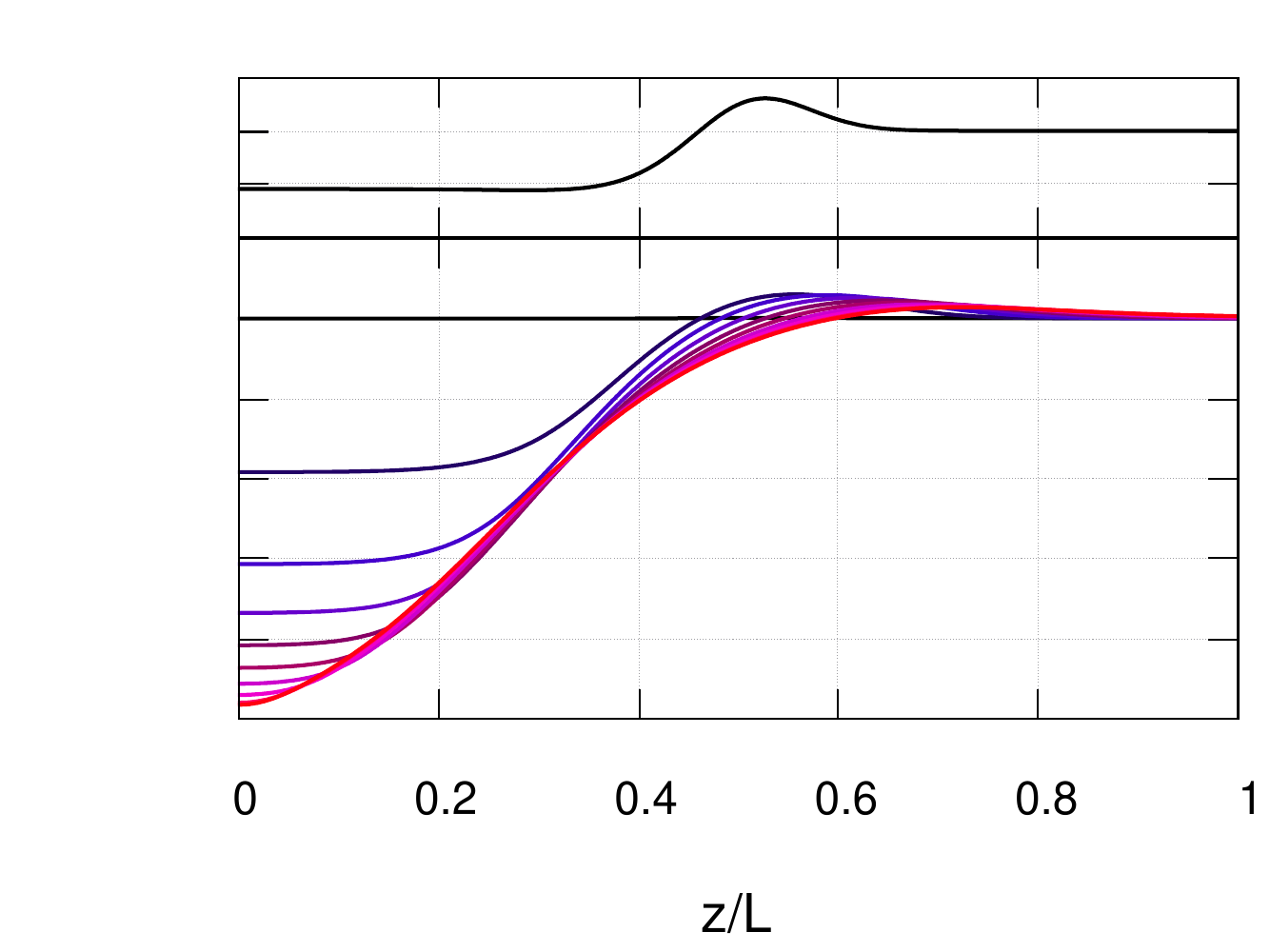}
        \subcaption{$\mu=1.2$}  
    \end{minipage}
    \hspace{-40pt}
    \begin{minipage}[b]{0.4\linewidth}
        \centering
        \includegraphics[width=1.\linewidth]{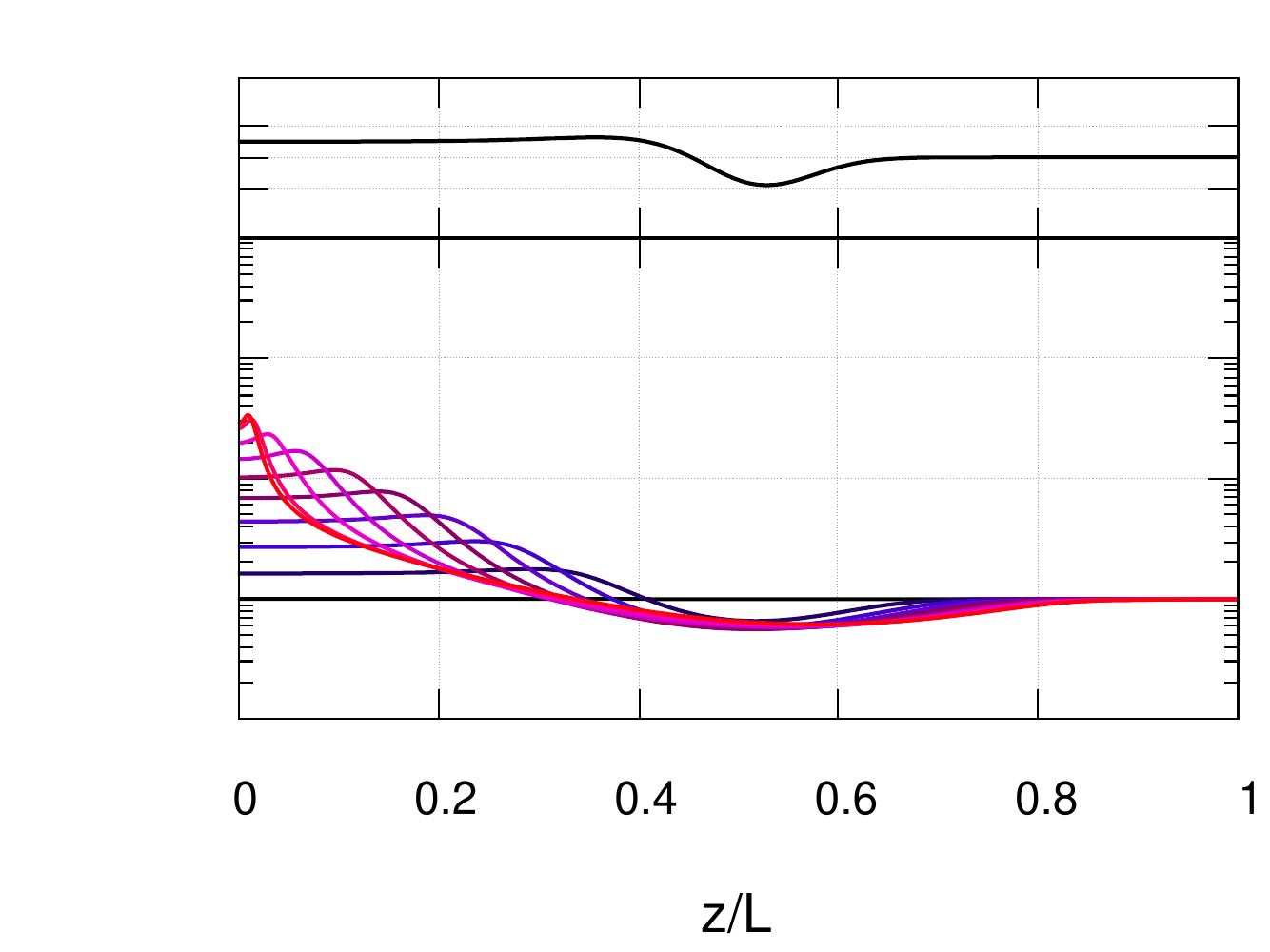}
        \includegraphics[width=1.\linewidth]{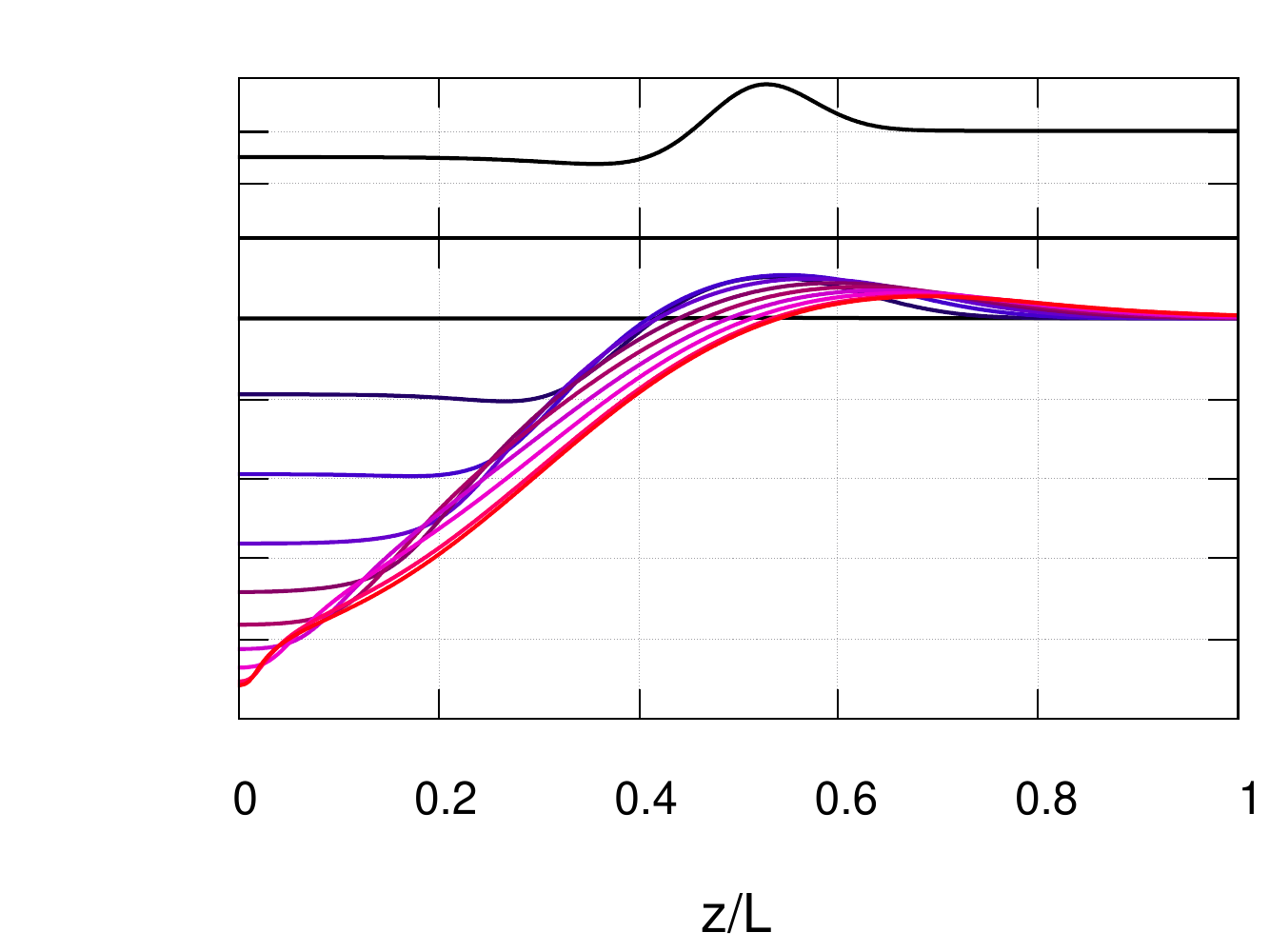}
        \subcaption{$\mu=1.8$}
    \end{minipage}
    \caption{Snapshots of the fluid energy density contrast $\rho/\rho_\text{bg}$ and lapse function $\alpha$ for $\zeta(r)$ given by Eq.~(\ref{eq:Gprofile}) for $\mu=0.5$, $1.2$ and $1.8$ in the left, middle and right panels, respectively. The initial configurations ($t = 0.01 t_H$) are shown at the top of each panel in black on a different scale. We can see that the evolution does not lead to black hole formation for $\mu=0.5$ but does for $\mu=1.2$ and $1.8$.}
    \label{fig:SnapshotsRhoRhobg}
\end{figure}

First, we show the time evolutions of the
fluid energy density and the lapse function in Fig.~\ref{fig:SnapshotsRhoRhobg} for $\mu=0.5$, $1.2$, and $1.8$.
For the cases of type I fluctuations with $\mu=0.5$ (left) and $\mu=1.2$ (middle), the former disperses, but the latter contracts and forms a peak at the center in the density plot in Fig.~\ref{fig:SnapshotsRhoRhobg}.
The latter case (middle) indicates a typical gravitational collapse, for which the lapse is decreasing to a very small value as plotted in Fig.~\ref{fig:SnapshotsRhoRhobg} as in Ref.~\cite{PhysRevD.105.103538}.
For a type II fluctuation case with $\mu=1.8$ (right), the collapsing behavior occurs similarly to the $\mu=1.2$ (middle) case in the density and lapse plots.

\begin{figure}[h]
    \centering
    \hspace{-40pt}
    \begin{minipage}{0.4\linewidth}
        \centering
        \includegraphics[width=1.\linewidth]{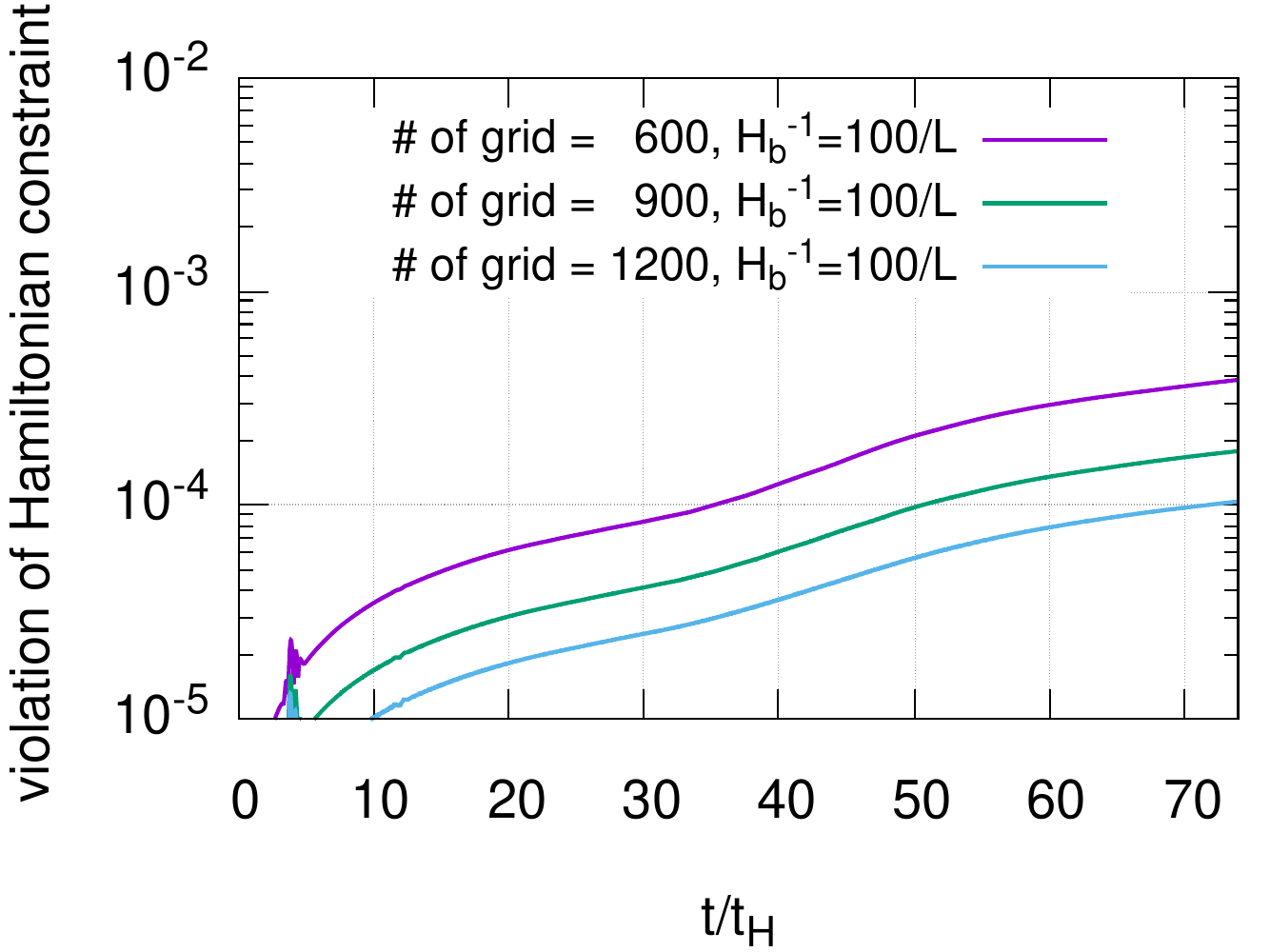}
        \subcaption{$\mu=0.5$}
    \end{minipage}
    \hspace{-40pt}
    \begin{minipage}{0.4\linewidth}
        \centering
        \includegraphics[width=1.\linewidth]{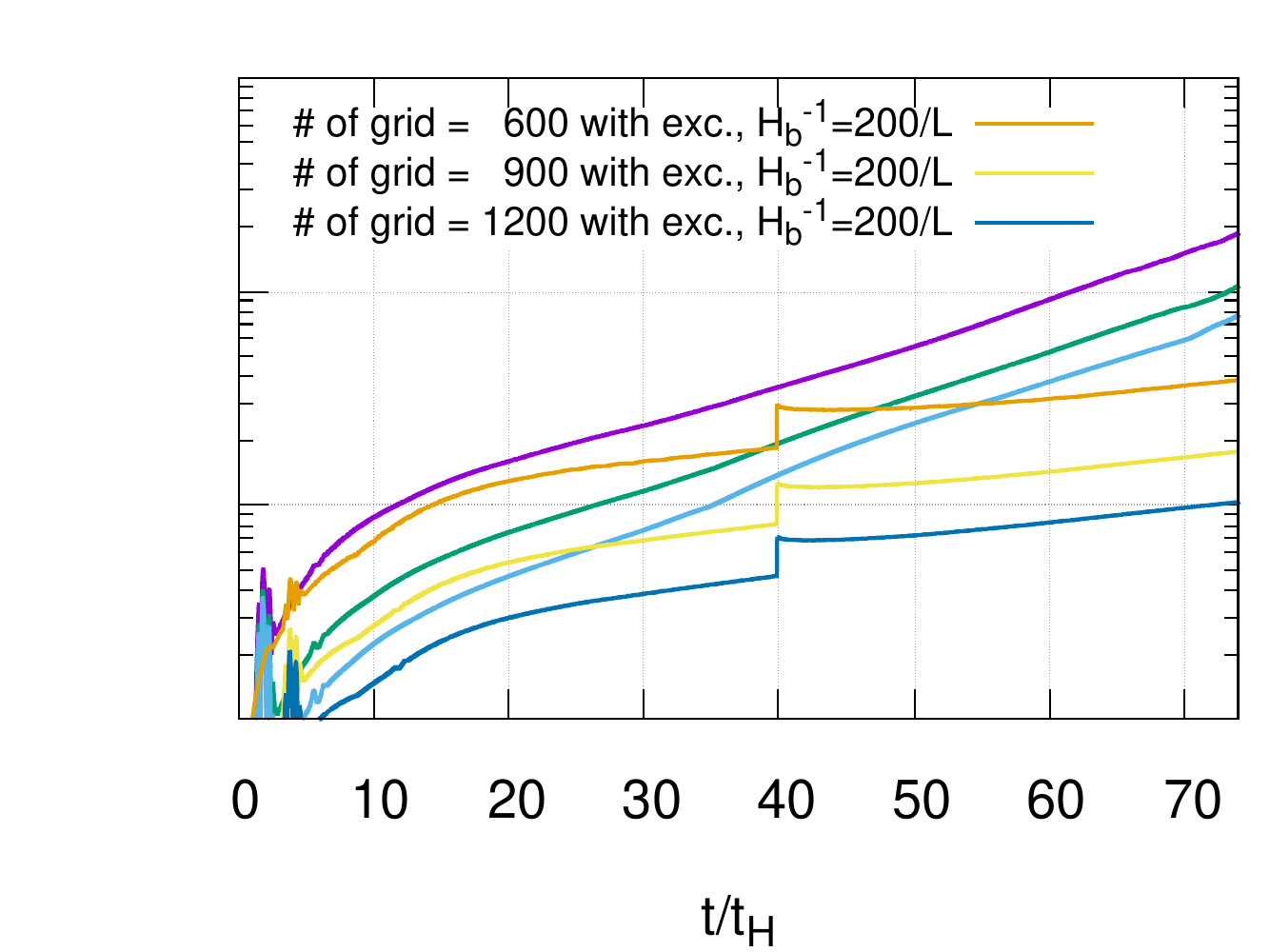}
        \subcaption{$\mu=1.2$}
    \end{minipage}
    \hspace{-40pt}
    \begin{minipage}{0.4\linewidth}
        \centering
        \includegraphics[width=1.\linewidth]{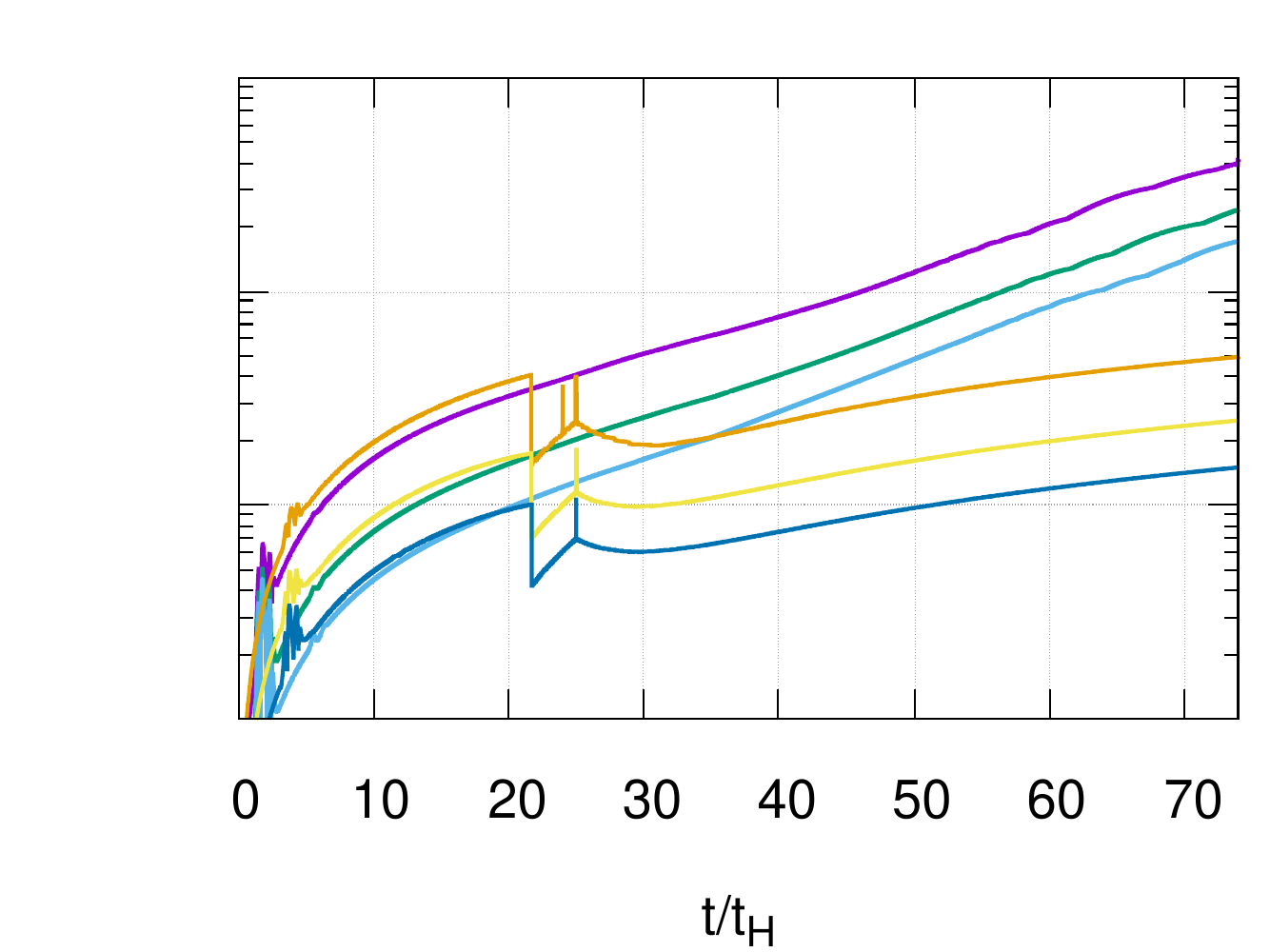}
        \subcaption{$\mu=1.8$}
    \end{minipage}
    \caption{
        The average value of the Hamiltonian constraint violation with different resolutions of the number of grid points for $\zeta(r)$ given by Eq.~(\ref{eq:Gprofile}) for $\mu=0.5$, $1.2$ and $1.8$ in the left, middle and right panels, respectively. 
        The horizontal axis is a time coordinate normalized by the time of the horizon entry.
        We use the simulation result upto $t=50t_H$ to deduce spacetime structures near the horizon formation before the violation propagation reaches the bulk region from the outer boundary. 
        For computing a longer time evolution of the PBH mass, the average is taken outside the BH apparent horizon (or the future-outer trapping horizon, described in Sec.~\ref{subsec:AHs}) after the horizon formation at around $t=40t_H$ for $\mu=1.2$ and $t=24t_H$ for $\mu=1.8$.
        }
    \label{fig:HamConstVio}
\end{figure}

To check the convergence of the simulation,
we plot the spatial average value of the Hamiltonian constraint\footnote{
    First, we evaluate the appropriately normalized value of the constraint violation $\mathcal H_p$ at each grid point labeled by $p$. Then the average is taken as $\sum_{p=1}^N \mathcal H_p/N$, where $N$ is the total number of the grid. 
}~\cite{Harada:2015yda} (Fig.~\ref{fig:HamConstVio}).
One can find roughly quadratic convergence in the different spatial resolutions, and it is consistent with our numerical code with a second-order scheme implemented in evaluating the fluid flux.

\subsection{Identification of BH apparent horizon}\label{subsec:AHs}
First, we introduce two independent future-directed radial null vector fields, $l$ and $l'$. 
We also define the null expansion for $l$ and $l'$ as $\theta$ and $\theta'$.
A null expansion is expressed as $\theta \coloneqq 2 \mathcal{L}_{l} R/R$, where $\mathcal{L}_l$ denotes the Lie derivative along $l$.
A trapping horizon is defined by a hypersurface foliated by marginal surfaces that satisfy $\theta\theta'=0$ or equivalently the compactness $2M/R=1$ with $M$ being the Misner--Sharp mass defined by 
\begin{equation}    
M(t,r) \coloneqq \frac{1}{2}R(1 - \nabla_\mu R\, \nabla^\mu R). 
\end{equation}
A trapping horizon is said to be future, past and bifurcating for $\theta+\theta'<0$, $\theta+\theta'>0$ and $\theta+\theta'=0$, respectively~\cite{PhysRev.136.B571,Hayward:1994bu,Hayward:1993wb,galaxies10060112,Maeda:2009tk}. 
A trapping horizon can be identified with an apparent horizon in the present setting. Black hole and cosmological apparent horizon are future and past trapping horizon, respectively.

\begin{figure}[h]
    \centering
    \hspace{-40pt}
    \begin{minipage}[b]{0.4\linewidth}
        \centering
        \includegraphics[width=\linewidth]{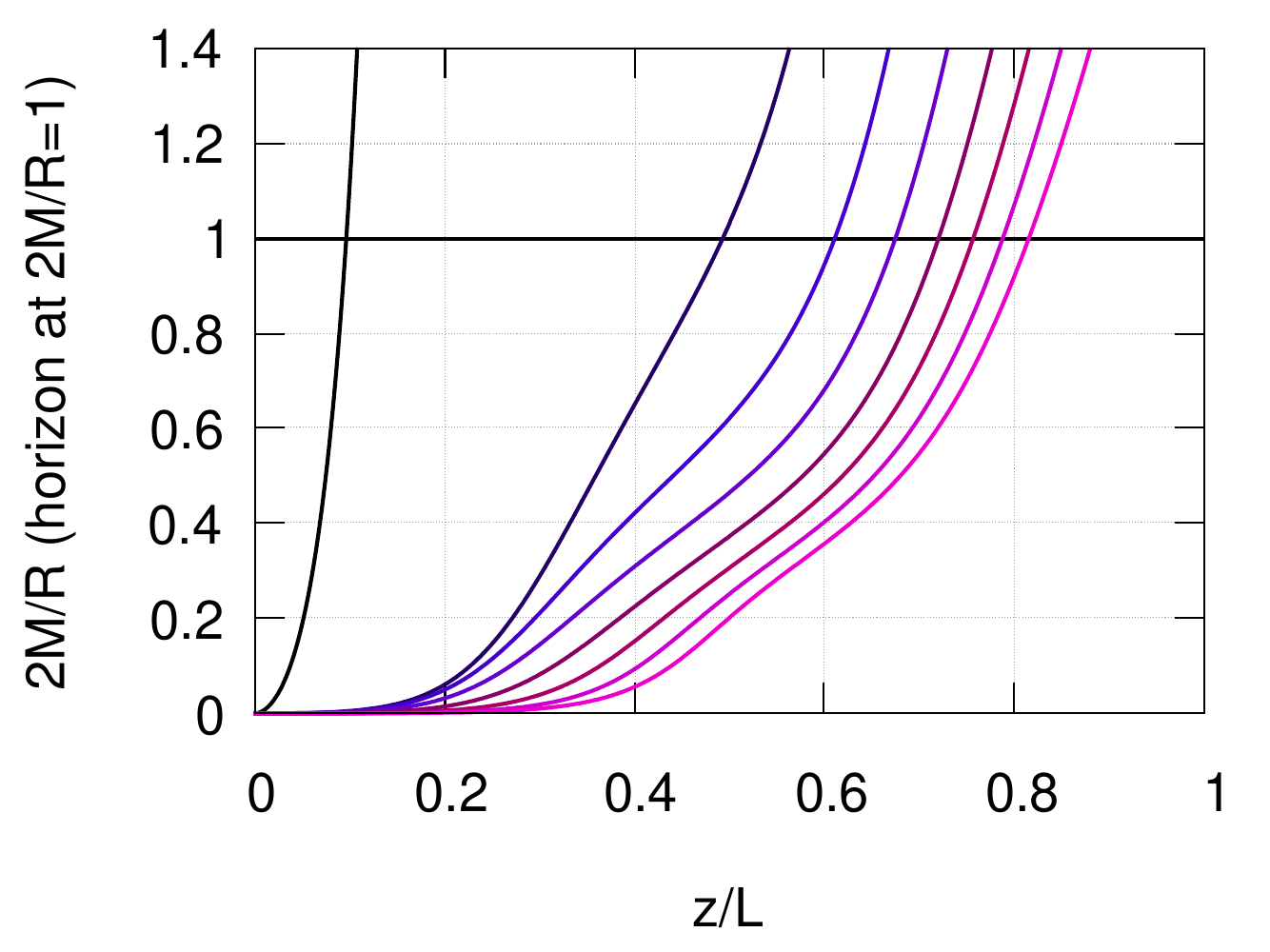}
        \subcaption{$\mu=0.5$}
    \end{minipage}
    \hspace{-40pt}
    \begin{minipage}[b]{0.4\linewidth}
        \centering
        \includegraphics[width=\linewidth]{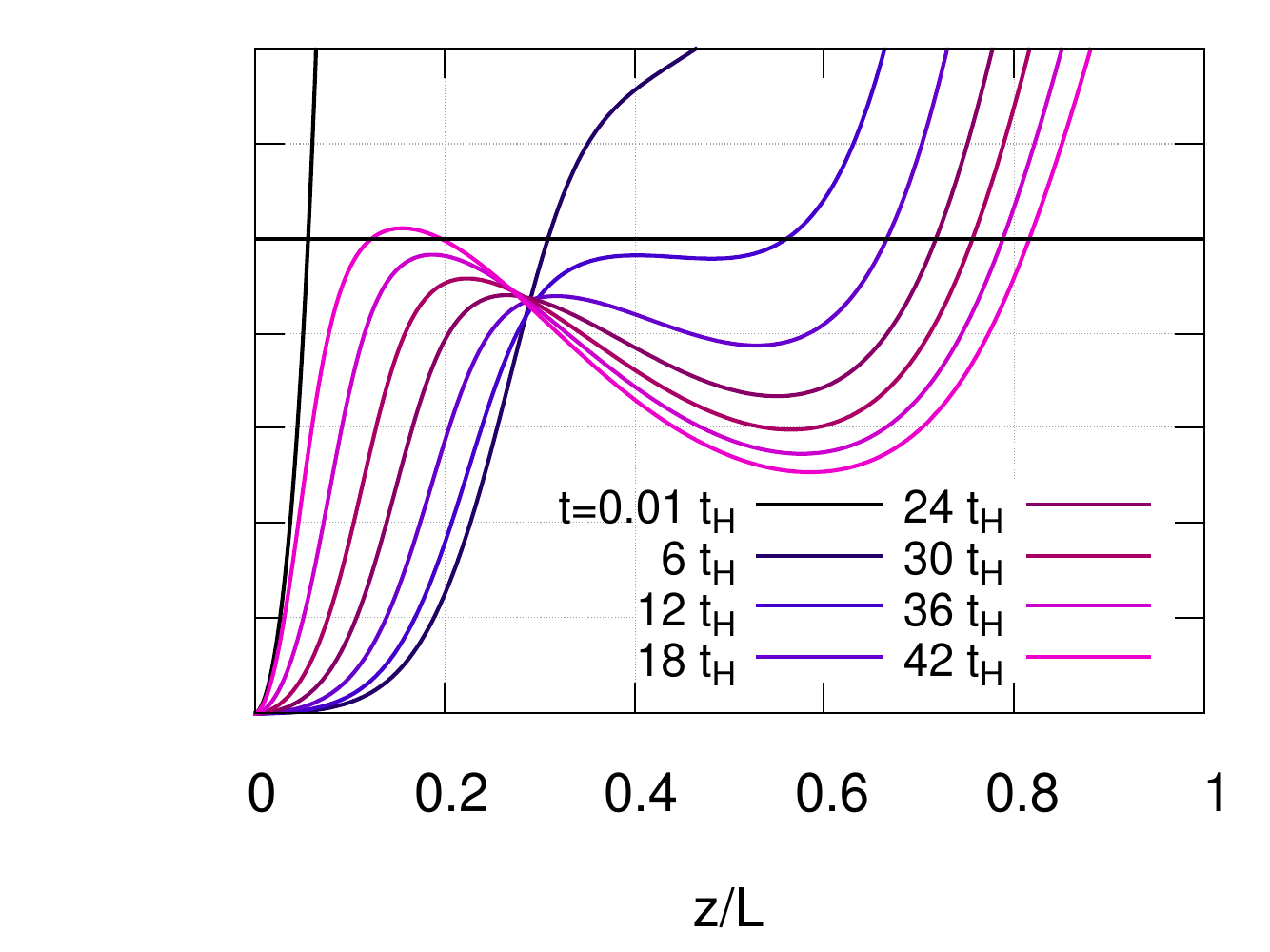}
        \subcaption{$\mu=1.2$}  
    \end{minipage}
    \hspace{-40pt}
    \begin{minipage}[b]{0.4\linewidth}
        \centering
        \includegraphics[width=\linewidth]{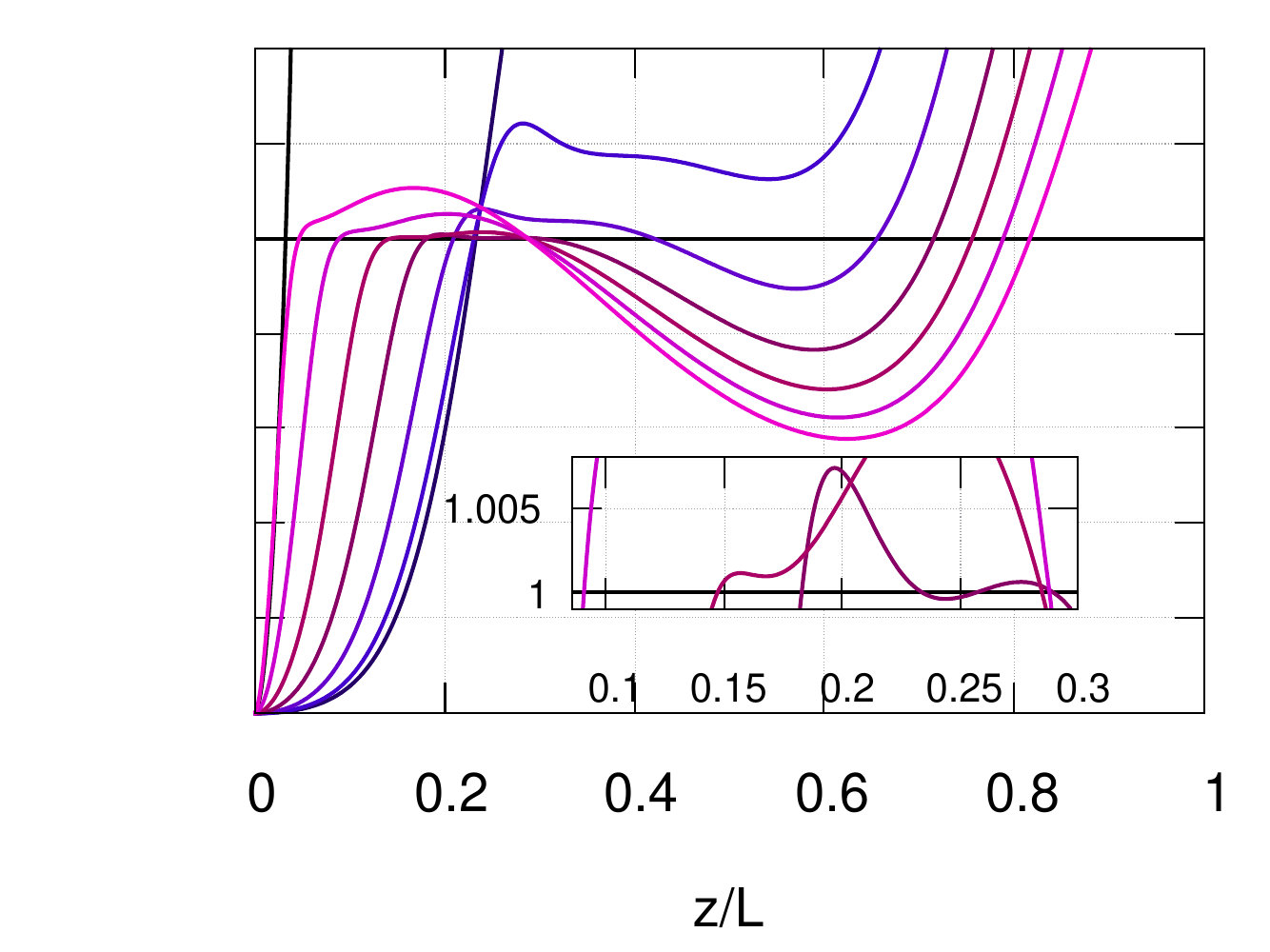}
        \subcaption{$\mu=1.8$}
    \end{minipage}
    \caption{
        Snapshots of the compactness $2M(t,r)/R(t,r)$ for $\zeta(r)$ given by Eq.~(\ref{eq:Gprofile}) for $\mu=0.5$, $1.2$ and $1.8$ in the left, middle and right panels, respectively.
        For all these three cases, we can see that at the earliest time $t=0.01 t_{H}$, the compactness monotonically increases from 0 at $z=0$ and crosses $1$ at some value of $z$, corresponding to a past trapping horizon or a cosmological horizon.
        For $\mu=0.5$, the compaction monotonically increases even in the later times and crosses $1$ only once even in the late times. 
        This implies that no black hole forms in this case.
        For $\mu=1.2$, after some evolution, the monotonicity breaks down. 
        It obtains a maximum and a minimum both smaller than $1$. The maximum is first smaller than $1$, becomes equal to $1$ at around $t=40 t_{H}$, and later gets larger than $1$, while the minimum continues to be below $1$. 
        As the maximum exceeds $1$, the number of intersection points changes as $1$, $2$, and $3$. 
        This implies the formation of a black hole trapping horizon inside the cosmological horizon at around $t=40 t_{H}$.
        For $\mu=1.8$, at first sight, the behavior is similar to that of $\mu=1.2$, but we can see that the maximum and minimum are larger than $1$ for $t=6t_H$ when the monotonicity breaks down. 
        As time proceeds further, the minimum decreases and gets smaller than $1$ for $t=12t_H$, while the maximum continues to be above $1$.
        Around $t=18t_H$, another set of maximum and minimum points appear between the pre-existing extrema, and the minimum gets smaller than 1 for $t=24t_H$.
        As the zoomed subplot shows, the number of intersection points becomes five for $t=24t_H$.
        Then, the outer maximum decreases to 1 and increases again after touching the horizontal line of $2M/R=1$. 
        As will be understood from Fig.~\ref{fig:AHg}, the future trapping horizon appears after this moment of the turnaround of the maximum. 
        After that, a similar behavior is observed for the inner maximum. Then, the minimum between the two maxima increases to a value larger than 1, and finally, there are 3 intersections with the line of $2M/R=1$.
    }
    \label{fig:compactness}
\end{figure}

\begin{figure}[h]
    \centering
    \hspace{-35pt}
    \begin{minipage}[b]{0.4\linewidth}
        \centering
        \includegraphics[width=\linewidth]{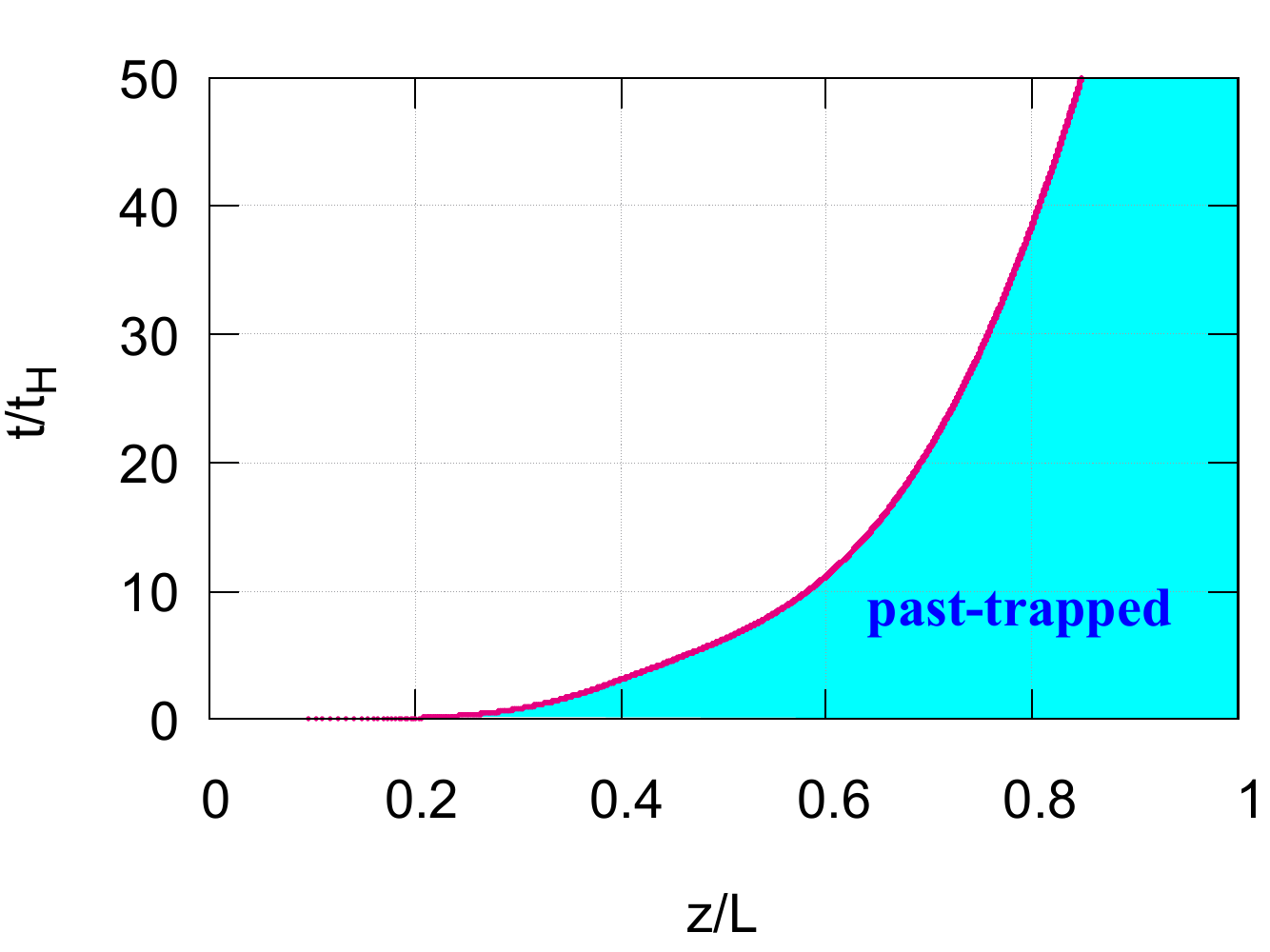}
        \subcaption{$\mu=0.5$}    
    \end{minipage}
    \hspace{-35pt}
    \begin{minipage}[b]{0.4\linewidth}
        \centering
        \includegraphics[width=\linewidth]{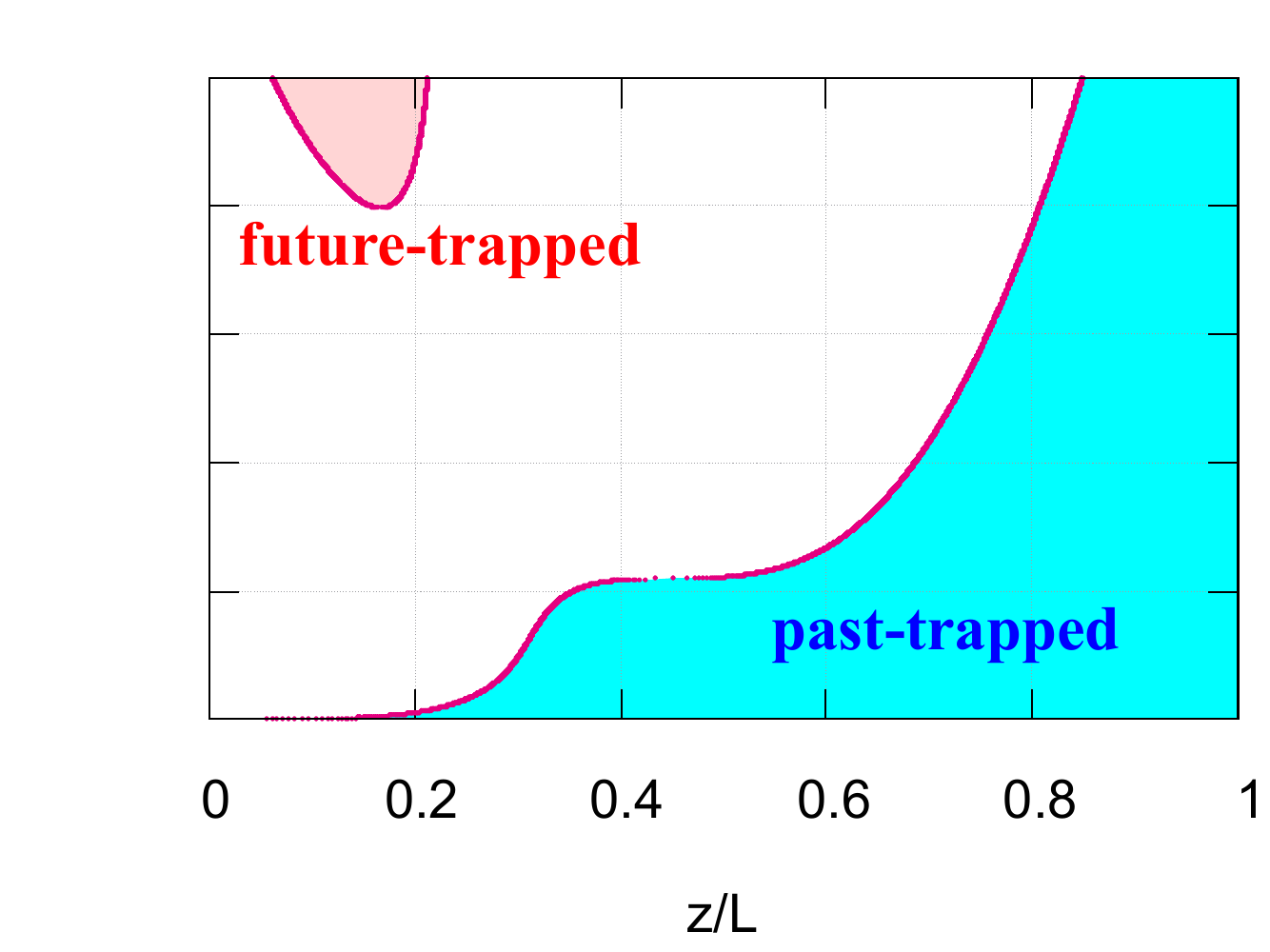}
        \subcaption{$\mu=1.2$}    
    \end{minipage}
    \hspace{-35pt}
    \begin{minipage}[b]{0.4\linewidth}
        \centering
        \includegraphics[width=\linewidth]{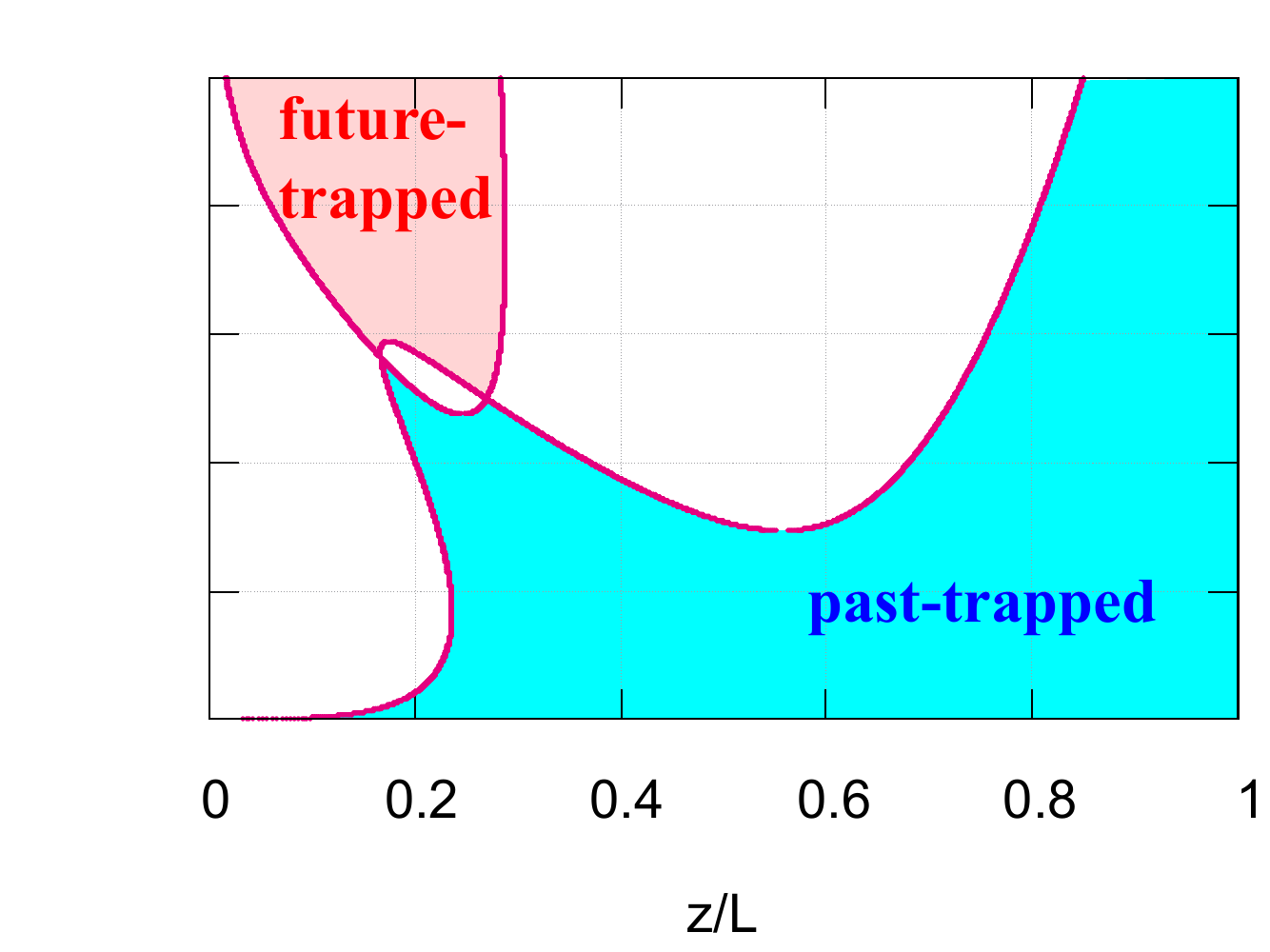}
        \subcaption{$\mu=1.8$}
    \end{minipage}
    \caption{
        Configurations of the trapping horizons $R(t,r)=2M(t,r)$ for $\zeta(r)$ given by Eq.~(\ref{eq:Gprofile}) for $\mu=0.5$, $1.2$ and $1.8$ in the left, middle and right panels, respectively.
        It corresponds to the points $(t,r)$ of the compactness $2M/R=1$ in Fig.~\ref{fig:compactness}.
        We can see that there is only a single past trapping horizon for $\mu=0.5$ and separate future trapping horizon and past trapping horizon for $\mu=1.2$. For $\mu=1.8$, two trapping horizons intersect each other twice. The intersection points correspond to bifurcating trapping horizons. The region surrounded by the trapping horizons is untrapped with the null expansions $\theta>0$ and $\theta'<0$. 
        The future and past boundaries of this region are future and past trapping horizons with the ends being bifurcating trapping horizons, respectively.
    }
    \label{fig:AHg}
\end{figure}

In Fig.~\ref{fig:compactness}, the snapshots of the compactness $\qty(2M/R)(z)$ are plotted for $\mu=0.5, 1.2$ and $1.8$.
For all these three cases, we can see that at the earliest time $t=0.01 t_{H}$, the compactness monotonically increases from 0 at $z=0$ and crosses $1$ at some value of $z$,  corresponding to a past trapping horizon or a cosmological horizon.
For $\mu=0.5$, the compaction monotonically increases even in the later times and crosses $1$ only once even in the late times.
This implies that no black hole forms in this case.
For $\mu=1.2$, after some evolution, the monotonicity breaks down. 
It obtains a maximum and a minimum both smaller than $1$. 
The maximum is first smaller than $1$ and later gets larger than $1$, while the minimum continues to be below $1$. 
As the maximum exceeds $1$, the number of intersection points changes as $1$, $2$, and $3$. 
This implies black hole formation inside the cosmological horizon.
For $\mu=1.8$, at first sight, the behavior is similar to that of $\mu=1.2$, but we can see that the maximum and minimum are larger than $1$ for $t=6t_H$, when the monotonicity breaks down. 
As time proceeds further, the minimum decreases and gets smaller than $1$ for $t=12t_H$, while the maximum continues to be above $1$.
Around $t=18t_H$, another set of maximum and minimum points appear between the pre-existing two extrema, and the minimum gets smaller than 1 for $t=24t_H$.
The number of intersection points becomes five.
Then the outer maximum decreases to 1 and increases again after touching the horizontal line of $2M/R=1$.
As will be understood from Fig.~\ref{fig:AHg}, the future trapping horizon appears after this moment of the turnaround of the maximum. 
After that, a similar behavior is observed for the inner maximum. Then the minimum between the two maxima increases to a value larger than 1, and finally, there are 3 intersections with the line of $2M/R=1$.

\begin{figure}[h]
    \centering
    \begin{minipage}[t]{.7\linewidth}
        \centering
        \includegraphics[width=\linewidth]{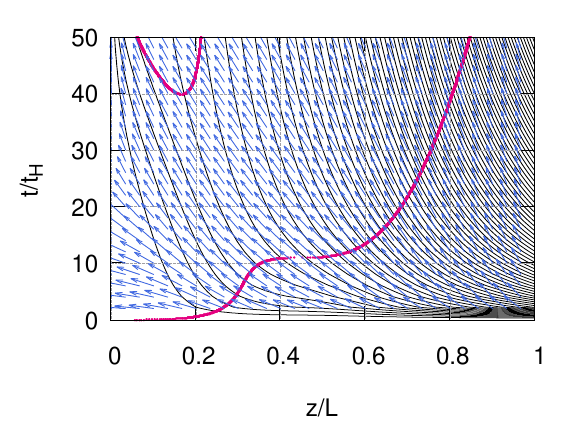}
    \end{minipage}
    
    \begin{minipage}[t]{.7\linewidth}
        \centering
        \includegraphics[width=\linewidth]{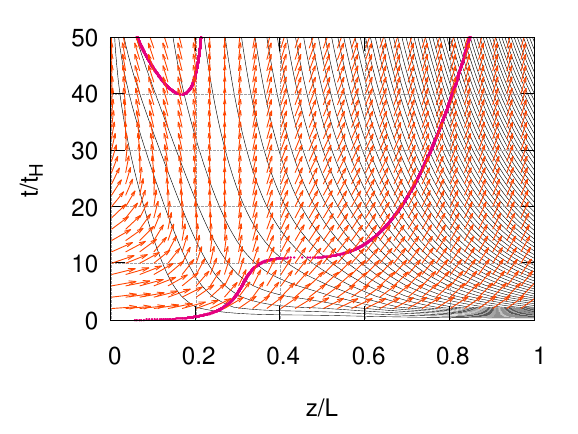}
    \end{minipage}

    \caption{
        Directions of two independent null vector fields $l$ and $l'$ are depicted with blue and orange arrows in the top and bottom panels, respectively, for $\zeta(r)$ given by Eq.~(\ref{eq:Gprofile}) for $\mu=1.2$.   
        The $R$ contour is plotted as black curves with 1.0 increments starting from $R=1.0$ and increasing from left bottom to right top.
        The red curves indicate trapping horizons satisfying $\theta\theta'=0$.
        Comparing the directions of the null vector fields and the contour curves of the areal radius, we can see that the red and blue shaded regions in Fig.~\ref{fig:AHg} are future- and past-trapped regions, respectively. 
        More specifically, in the future (past)-trapped region, the areal radius decreases (increases) along both of the arrows.
    }
    \label{fig:typeInull}
\end{figure}

\begin{figure}[h]
    \centering
    \begin{minipage}[t]{\linewidth}
        \centering
        \includegraphics[width=.7\linewidth]{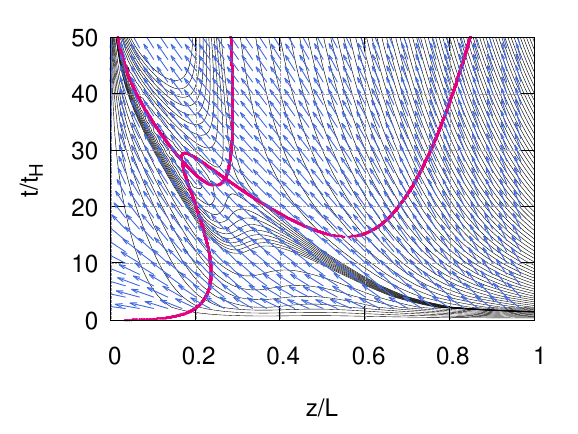}
    \end{minipage}
    
    \begin{minipage}[t]{\linewidth}
        \centering
        \includegraphics[width=.7\linewidth]{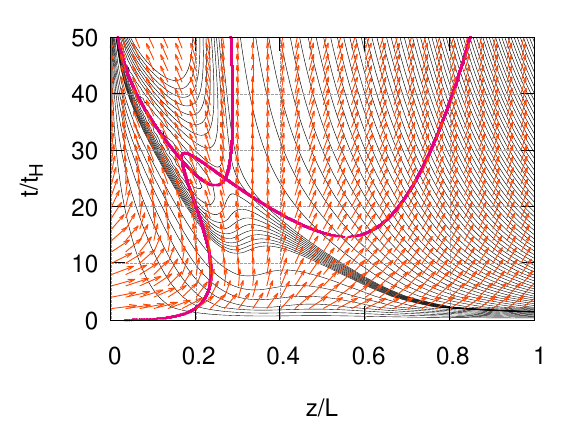}
    \end{minipage}
    \caption{
        Same as Fig.~\ref{fig:typeInull} but for $\mu=1.8$. 
        The $R$ contour is plotted as black curves with 1.0 increments starting from $R=1.0$ and increasing from left bottom to right top.
        In addition to the original plot of $R$ levels with increments of 1.0, it is also plotted with increments of 0.1 from $R= 8.1$ to $8.9$.
        We can see that two trapping horizons intersect each other twice. The intersection points correspond to bifurcating trapping horizons. Looking at these panels very carefully, we can see that the intersection points on the left-hand and right-hand sides are a maximum and a saddle, respectively, in terms of the areal radius. The region surrounded by the trapping horizons is untrapped with the null expansions $\theta>0$ and $\theta'<0$. The future and past boundaries of this region are future and past trapping horizons with the ends being bifurcating trapping horizons, respectively. The enlarged views around the intersection points by the two red curves and the region surrounded by the red curves are provided in Fig.~\ref{fig:typeIInullLarge}.
    }
    \label{fig:typeIInull}
\end{figure}

The worldlines of the intersection points, which correspond to trapping horizons, are shown in Fig.~\ref{fig:AHg}.
We can see a single trapping horizon for $\mu=0.5$, 
two distinct trapping horizons for $\mu=1.2$ and two intersecting trapping horizons for $\mu=1.8$.
To identify future-trapped ($\theta<0$ and $\theta'<0$) and past-trapped ($\theta>0$ and $\theta'>0$) regions, we also show directions of two independent null vector fields $l$ and $l'$ together with the contour curves of the areal radius in Figs.~\ref{fig:typeInull} and \ref{fig:typeIInull}. 
Comparing the directions of the null vector fields and the contour curves of the areal radius, we can realize that the red and blue shaded regions in Fig.~\ref{fig:AHg} are future- and past-trapped regions, respectively. 
More specifically, in the future (past)-trapped region, the areal radius decreases (increases) along both of the arrows, and the red curve indicates the boundary satisfying $\theta\theta'=0$.  
Then, the boundary of the future (past)-trapped region is a future (past) trapping horizon. We can see in Fig.~\ref{fig:AHg} that there is only a single past trapping horizon for $\mu=0.5$ and separate future trapping horizon and past trapping horizon for $\mu=1.2$. For $\mu=1.8$, two trapping horizons intersect each other twice. The intersection points correspond to bifurcating trapping horizons. The enlarged views of Fig.~\ref{fig:typeIInull} around the intersection points and the region surrounded by the red curves are provided in Fig.~\ref{fig:typeIInullLarge}.

It is not easy to understand the horizon configuration, in particular, for the case of $\mu=1.8$. 
Therefore, we carefully investigate it and present a more concrete classification of trapping horizons in the following sections. 
Readers who would like to get a quick intuition about the horizon configuration can refer to the schematic conformal diagram in Figs. \ref{fig:diagrams2} and \ref{fig:PBHponchiDust}.


\begin{figure}[h]
    \centering
    \begin{minipage}[t]{\linewidth}
        \centering
        \fbox{\includegraphics[width=.5\linewidth,trim= 75 105 150 65, clip]{fig/CHNin_mu1.8.pdf}}
    \end{minipage}
    
    \begin{minipage}[t]{\linewidth}
        \centering
        \fbox{\includegraphics[width=.5\linewidth,trim= 75 105 150 65, clip]{fig/CHNout_mu1.8.pdf}}
    \end{minipage}
    \caption{Same as Fig.~\ref{fig:typeIInull} but the enlarged views of the complicated region near the intersection points.  Note that the red curves divide the whole region into five, i.e., the bottom left (I), top middle (II), top right (III), bottom right (IV), and the region in the center of the panel surrounded by the red curves (V). 
    On each panel, $R$ is small on the bottom left, top middle, and bottom right, while large on the top right. 
    So, we can conclude that the left-hand and right-hand intersection points of the red curves correspond to a maximum and a saddle, respectively, in terms of $R$. We can see that the blue arrows on the top panel are tangent to the $R$-contour on the red curve convex upward, implying that $\theta=0$ there, while the orange arrows on the bottom panel are tangent to the $R$-contour on the red curve convex downward, implying that $\theta'=0$.
    Thus, the signs of the null expansions $(\theta,\theta')$ in regions I, II, III, and IV are given by $(-,+)$, $(-,-)$, $(-,+)$ and $(+,+)$, respectively. We can see that the signs are $(+,-)$ in region V, implying that this region is untrapped. 
    The part of the red curve convex upward that is the boundary of region V has the signs $(0,-)$, implying that it is a future trapping horizon.  
    The same analysis implies that the part of the red curve convex downward that is also the boundary of region V is a past trapping horizon.
    } 
    \label{fig:typeIInullLarge}
\end{figure}

\subsection{Spacetime structures of type A/B PBH}\label{subsec:3zonePic}
Let us introduce the classification of trapping horizons following Refs.~\cite{Hayward:1993wb,Maeda:2009tk,Harada_2018}. 
For this purpose, to avoid confusion, we introduce another notation for the two radial null vector fields and associated expansions as $l_\pm$ and $\theta_\pm$ for a specific trapping horizon.
Then we always set $\theta_+=0$ when considering a trapping horizon satisfying $\theta_+\theta_-=0$. 
The classification of the trapping horizons can be summarized in Table \ref{table:table_FOTH}.
\begin{table}[h]
    \centering
    \begin{tabular}{c|ccc}
        $\theta_+=0$            & +     & 0           & -   \\ \hline 
        $\theta_-$              & Past  & Bifurcating & Future \\ 
        $\mathcal{L}_-\theta_+$ & Inner & Degenerate  & Outer \\ 
    \end{tabular}
    \caption{
        Classification of trapping horizons satisfying $\theta_+\theta_-=0$ \cite{Hayward:1993wb}. 
        A black hole, a white hole, and cosmological trapping horizons with $w=1/3$ $(w=0)$ are identified with future-outer, past-outer, and past-degenerate (past-inner) trapping horizons, respectively.
    }
    \label{table:table_FOTH}
\end{table}

\begin{figure}[h]
    \centering
    \begin{minipage}{0.32\linewidth}
        \centering
        \includegraphics[width=\linewidth,page=1]{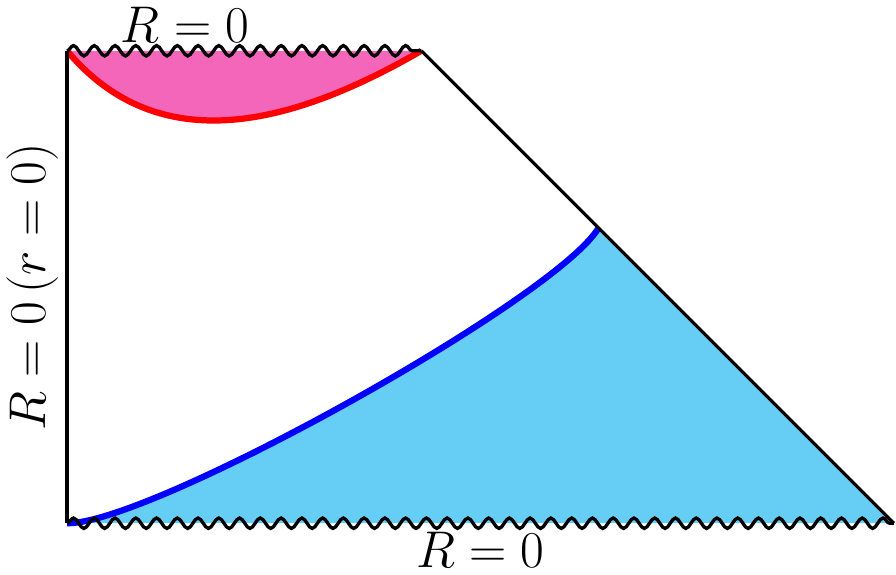}
        \subcaption{Type A PBH}
        \label{fig:PBHponchiI}
    \end{minipage}
    \begin{minipage}{0.64\linewidth}
        \centering 
        \includegraphics[width=.9\linewidth,page=2]{fig/fig1to11_04241402.pdf}
        \subcaption{Type B PBH}
        \label{fig:PBHponchiII}
    \end{minipage}
    \caption{
    The conformal diagrams inferred from the present numerical results in radiation domination.  The left panel defines a type A PBH as a causal structure of the typical case of PBH.
        The right one includes bifurcating trapping horizons and defines a type B PBH. 
        The red and blue curves depict the future and past trapping horizons, respectively. 
        The bifurcating trapping horizons ($\theta_+=\theta_-=0$) are depicted as circles in the type B case. 
    }
    \label{fig:PBHponchi}
\end{figure}

Unlike the dust case discussed in Appendix~\ref{sec:3zonePonchi}, we can only obtain numerical solutions for radiation case, such as those shown in Figs.~\ref{fig:typeInull} and \ref{fig:typeIInull} for $\mu=1.2$ and $1.8$, respectively.
The trajectories of $\theta=0$ and $\theta'=0$ are independently given by two smooth curves indicated by red lines in the top and bottom panels, respectively. 
In the top panels, on the smooth red line in the bottom right, the blue arrows are tangent to the solid black lines indicating the contour of the areal radius, which implies $\theta=0$ there.
In the bottom panels, on the other hand, the orange arrows are tangent to the solid black lines on the other smooth red line that is convex downward in the top left, which implies $\theta'=0$ there.
A very careful look at Figs.~\ref{fig:typeIInull} and \ref{fig:typeIInullLarge} reveals that the intersection points on the left-hand and right-hand sides are a maximum and a saddle, respectively, in terms of the areal radius.
From these figures, we can find that the region surrounded by the trapping horizons, which we call region V, is untrapped with the null expansions $\theta>0$ and $\theta'<0$. 
The part of the red curve convex upward that is the boundary of region V is a future trapping horizon, while the part of the red curve convex downward that is the boundary of region V is a past trapping horizon.

From these figures and carefully comparing the similarities and differences from the exact conformal diagrams of the dust solution discussed in the Appendix~\ref{sec:3zonePonchi}, we can deduce possible conformal diagrams and proper identifications of the trapping horizons for the radiation case as shown in the left and right panels in Fig.~\ref{fig:PBHponchi} as possible conformal diagrams of type A and type B PBHs, respectively. 

The crucial difference between the right and left panels in Fig.~\ref{fig:PBHponchi} is the existence of the bifurcating trapping horizons $(\theta_-;\mathcal{L}_-\theta_+)|_{\theta_+=0}=(0;*)$, which is described by the intersection points of the two trapping horizon trajectories in Fig.~\ref{fig:typeIInull}. 
In this paper, we propose to use this difference to classify the PBH formation processes into types A and B; that is, we identify the PBH as type A if there is no bifurcating trapping horizon and as type B if at least one bifurcating trapping horizon appears during the time evolution associated with the intersection point. 
KHW has already pointed out this difference in Ref.~\cite{PhysRevD.83.124025} for the case of the dust fluid. 
Unlike the current proposal, the authors in Ref.~\cite{PhysRevD.83.124025} did not distinguish between the definitions of type I/II and type A/B.
However, in our simulation with $w=1/3$, we find the case of type II-A, the horizon configuration is of type A from type II fluctuation for the $\mu=1.4$ case. 
This fact is our motivation for the new definition of the PBH.

The critical difference between the dust and the radiation cases is that trapping horizons in the FLRW regions are timelike and null for the former and the latter, respectively. 
See, e.g.,  Ref.~\cite{Harada_2018}. 
Then, we confirm that the outermost boundary surface of the future-trapped region is the future-outer trapping horizon, which describes a BH apparent horizon. 
Therefore, the outermost boundary surface between the past-trapped and untrapped regions, which describes a cosmological apparent horizon, asymptotically approaches the past degenerate trapping horizon in the background FLRW universe.

It is worth noting that the schematic diagram given by Fig.~\ref{fig:PBHponchiII} implies that any light ray emitted from the center reaches not the future null infinity but the future singularity of the type B PBH.
Although we cannot make any general statement for the moment, the null trajectories indicated by the orange arrows in Fig.~\ref{fig:typeIInull} seem consistent with this observation.

\section{PBH mass}\label{subsec:PBHmass}
\subsection{PBH mass with the Gaussian initial fluctuation}
\label{subsec:PBHmass_Gaus}
\begin{figure}[h]
    \centering
    \includegraphics[width=.7\linewidth]{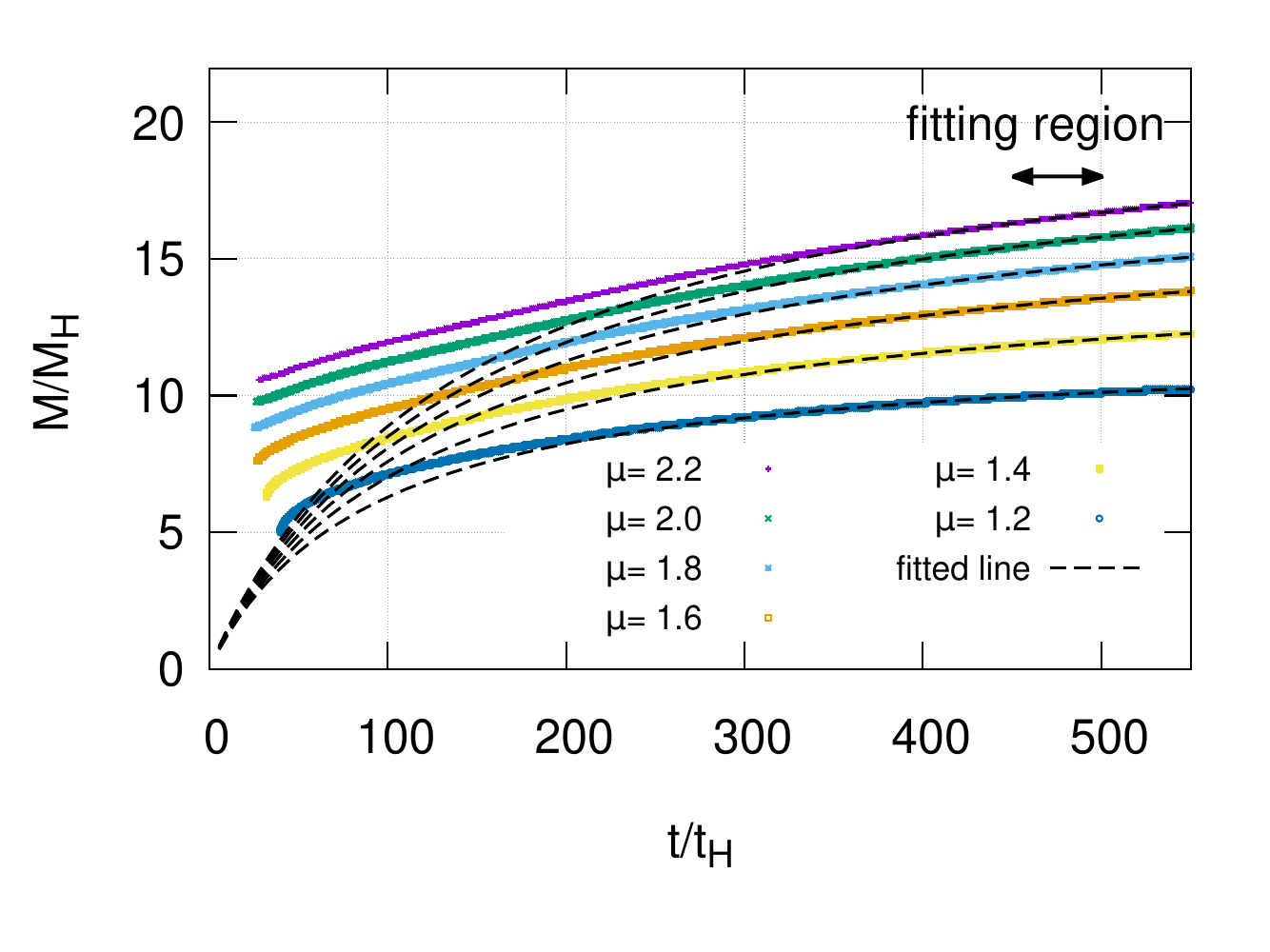}
    \caption{
        The time evolution of the BH mass for $\zeta(r)$ given by Eq.~(\ref{eq:Gprofile}) for $\mu=1.2, 1.4,1.6,1.8,2.0$ and $2.2$.
        The black dashed curves are fitting curves with Eq.~\eqref{eq:Mt} including the two parameters $M_f$ and $F$.
        The fitting region is taken for the interval between $t=450t_H$ and $t=500t_H$. 
        The values of fitting parameters are shown in Table~\ref{table:table}.
    }
    \label{fig:evolPBHmass}
\end{figure}

\begin{figure}[h]
    \centering
       \includegraphics[width=.8\linewidth]{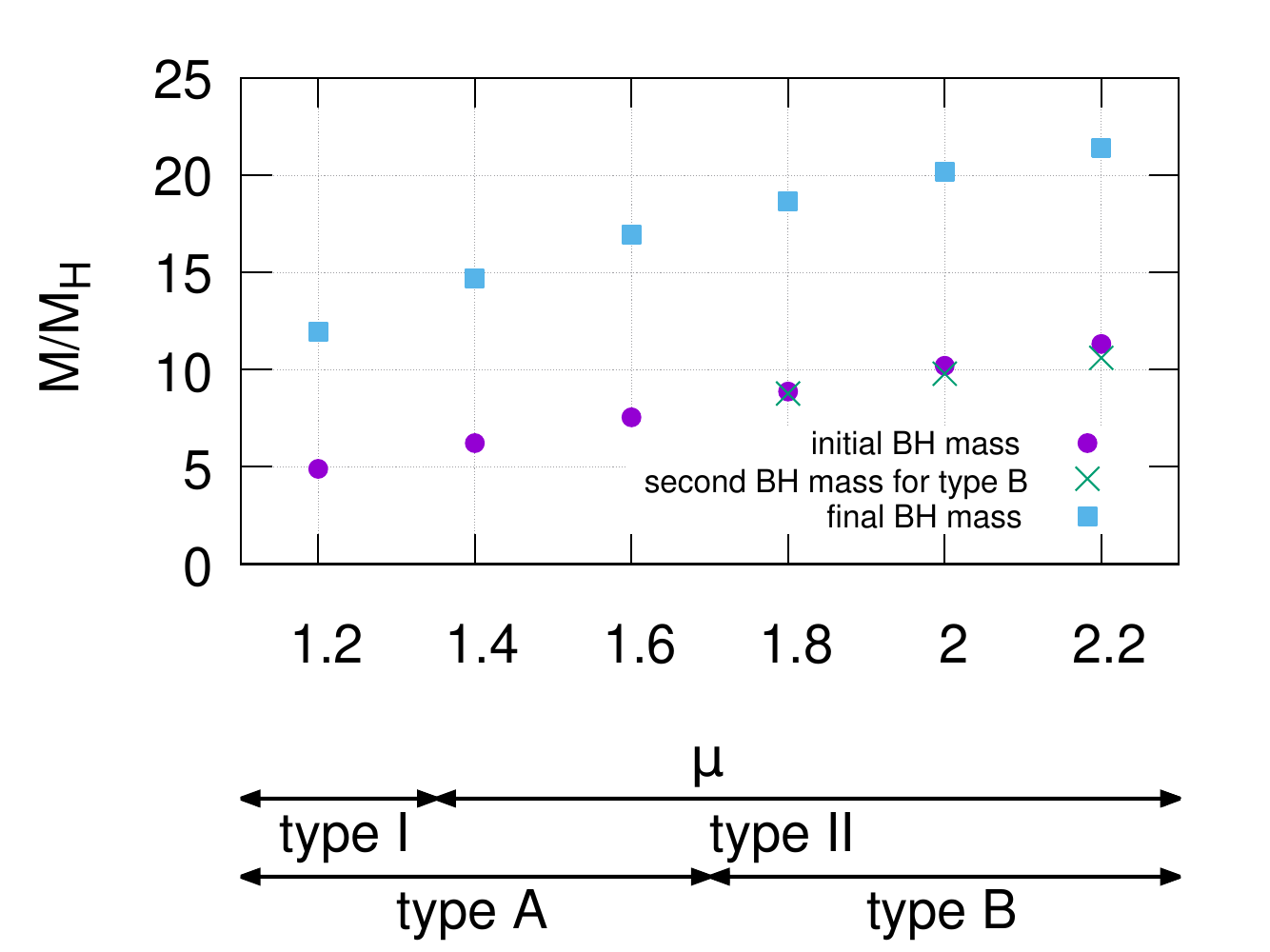}
    \caption{
        PBH mass in terms of the horizon mass at horizon entry $M_H$ to the amplitude of fluctuation $\mu$ for $\zeta(r)$ given by Eq.~(\ref{eq:Gprofile}).
        In this case, in terms of fluctuation types, $\mu = 1.2$ is classified as type I, while $\mu=1.4 \:\text{to}\: 2.2$ as type II, as seen in  Fig. \ref{fig:compfunc}.
        In terms of formation types, on the other hand, $\mu = 1.2$ to $1.6$ are classified as type A, while $\mu=1.8$ to $2.2$ as type B, as seen in Figs. \ref{fig:AHg} and \ref{fig:PBHponchi}. 
        We can see that the initial and final PBH masses get larger with increasing amplitude $\mu$.
    }
    \label{fig:mass_amp}
\end{figure}

Let us introduce the mass of the PBH in two different ways at each characteristic time.
The first is to calculate the Misner-Sharp mass at the apparent horizon at its first appearance with the future-trapped region (see Fig.~\ref{fig:AHg}). 
We refer to the mass at this time as the initial BH mass. 
In a type A PBH, the BH apparent horizon appears as a future-outer trapping horizon, while it appears as a bifurcating outer trapping horizon in type B. 
We also calculate the final PBH mass by estimating the effect of the accretion. 
The accretion process is well described by the mass accretion model, often called Novikov-Zeldovich or Bondi accretion~\cite{Zeldovich:1967lct,1952MNRAS.112..195B}. 
In this accretion model, the time evolution of the mass is given by 
\begin{equation}
    \dv{M}{t}=16\pi F M^2 \rho_\text{b},
    \label{eq:NZ_formula}
\end{equation}
where $\rho_b = \rho_{b,0} (t_0/t)^2$ is the background energy density, and $F$ is a nondimensional constant that corresponds to the efficiency of accretion, which is commonly numerically found to be of order $\mathcal{O}(1)$ and will depend on the details of the accretion process.
It is important to note that Eq.~\eqref{eq:NZ_formula} is not valid at the time of formation of the trapping horizon since it neglects the cosmological expansion as Carr and Hawking (1974)~\cite{10.1093/mnras/168.2.399} already pointed out. 
However, once we have a quasi-stationary flow into the PBH, it will correctly describe the accretion into the PBHs for sufficiently long times.
The integrated solution is,
\begin{equation}
         M(t)=\frac{1}{\frac{1}{M_f}+\frac{3F}{2t}},
         \label{eq:Mt}
     \end{equation}
where $M_f$ is the final BH mass. 
We find the values of $M_f$ and $F$ by fitting Eq.~\eqref{eq:Mt} to the numerical data in a sufficiently late finite time interval denoted in Fig.~\ref{fig:evolPBHmass} together with the time evolution of the mass for each value of $\mu$. 
The values of $F$ obtained by the fitting are listed in Table \ref{table:table}. 
The value of $F$ decreases with increasing $\mu$, which means the accretion is getting smaller. 
This behavior is consistent with the results in Ref.~\cite{escriva_2021}.
\begin{table}[H]
    \centering
    \begin{tabular}{l|cc}
        $\mu$   &   $M_{f}/M_{\text{H}}$    &   $F$\\ \hline 
        1.2     &   11.9  &   5.0 \\
        1.4     &   14.7  &   5.0 \\
        1.6     &   16.8  &   4.9 \\
        1.8     &   18.6  &   4.7 \\
        2.0     &   20.1  &   4.5 \\
        2.2     &   21.4  &   4.4 \\ 
    \end{tabular}
    \caption{
        The values of $F$ and $M_f$ for $\zeta(r)$ given by Eq.~(\ref{eq:Gprofile}) for $\mu=1.2, 1.4,1.6,1.8,2.0$ and $2.2$ obtained from the fitting in Fig.~\ref{fig:evolPBHmass}.
        }
    \label{table:table}
\end{table}
Let us check the $\mu$ dependence of the PBH mass.
In Fig.~\ref{fig:mass_amp}, we plot the initial and final PBH mass values for each value of $\mu$. 
This behavior can be seen for an amplitude much larger than the critical regime, where $M$ obeys a power law against $(\mu-\mu_{c})$~\cite{PhysRevLett.70.9,Evans:1994pj,Koike:1995jm,Niemeyer:1999ak,Musco:2012au}.
The initial and final PBH masses get larger with increasing amplitude $\mu$.
We will check whether this behavior appears for a different profile or not in the next subsection.

\subsection{PBH mass with the initial fluctuation motivated by the three-zone model}\label{sec:profile_dep}
Readers may naively expect that the PBH mass is a monotonically increasing function of $\mu$ as is indeed confirmed for our specific initial Gaussian-shaped profile~\eqref{eq:Gprofile} in Sec.~\ref{subsec:PBHmass_Gaus}. However, careful consideration with the help of the three-zone model suggests a different conclusion.\footnote{
Here, we consider how the mass of the PBH depends on the initial amplitude $\mu$ of the fluctuation in the type II regime by using the three-zone model with dust as seen in Fig.~\ref{fig:PBHponchiDust}. 
The mass parameter gives the PBH mass for the Schwarzschild spacetime in the middle patch. 
Let us fix the scale of the spatial curvature of the closed FLRW region. The horizon area of the Schwarzschild region is fixed by the horizon area on the matching surface between the closed FLRW and the Schwarzschild regions. 
By increasing the initial amplitude, we expect the closed FLRW region to be extended and the boundary to be shifted more to the right. 
In the closed FLRW side, the boundary shift to the right decreases the area of the horizons. 
Therefore, the mass parameter for the Schwarzschild region is also expected to decrease. 
From the above consideration of the model with dust, we may naively guess that the mass of the resultant PBH would decrease with the initial amplitude value for the type II regime. 
In fact, for general equations of state, the Misner-Sharp mass of the truncated closed FLRW universe is given by 4$\pi \rho R^{3}/3$ with $R$ being the areal radius of the outer boundary of the region, thus decreasing if the closed FLRW universe goes beyond the great sphere and approaches the ``separate universe scale'', as indicated in Ref.~\cite{PhysRevD.91.084048}. 
However, the assumption of the three-zone model is not necessarily valid for the radiation case, and the behavior also strongly depends on the profile of the initial fluctuation.  
}
This subsection explicitly shows that the PBH mass can have a non-monotonic behavior as a function of $\mu$ depending on the initial profile of the curvature fluctuation. 

To construct a specific initial profile that can be compared with the former Gaussian profile as Eq.~\eqref{eq:Gprofile} and Fig.~\ref{fig:InitialGaussianProfile}, let us try to utilize the three-zone profile as a reference, an extension of the model used in Ref.~\cite{PhysRevD.83.124025}. 
Let us consider the closed FLRW region for the initial overdense region to realize a situation similar to the schematic conformal diagram. 
Then, as is shown in the Appendix~\ref{sec:zeta_KHW_derivation}, under the boundary condition $\zeta=0$ at the surface of the overdense region, we obtain the following form of $\zeta(r)$:
\begin{equation}\label{eq:cFLRWprofile}
    \zeta_\text{cFLRW}(r) = 2\ln \frac{\cos k \chi_a/2}{\sqrt{\cos^4  k \chi_a/2 + \frac{1}{4}k^2 r^2}}\quad \text{for}\quad r\leq r_a:=\frac{1}{k}\sin(k\chi_a), 
\end{equation}
where $k^{-1}$ is the comoving radius of $S^3$ associated with the closed FLRW and $0<\chi_a\leq \pi /k$ describes the radial coordinate of the edge of the closed FLRW region. 
Although this expression is relevant for the region $r\leq r_a$, this functional form cannot be directly used as the initial profile for our numerical simulation. 
This is because, in our simulation, we suppose that the initial profile is continuously connected to the flat FLRW background, namely, $\zeta(r)$ should vanish for $r>L$. 
To realize this condition, first, we simply extend the expression \eqref{eq:cFLRWprofile} to the region $r>r_a$, and use the following specific functional form of $\zeta(r)$ for the initial condition of our simulation: 
\begin{align}
    \zeta(r) &= 2 \tilde{\mu}\ln \qty(\Psi_3(r)-\Psi_3(L))W(r), \label{eq:zetacFLRW}\\
    \Psi_3(r) &\coloneqq  e^{\zeta_\text{cFLRW}(r)/2} + e^{\zeta_\text{ref}(r)/2}, \label{eq:psi3}
\end{align}
where $\zeta_\text{ref}\coloneqq 0.001 e^{-\frac{1}{2}k^2 r^2}$ and $\tilde{\mu} \coloneqq k \chi_a/\pi$. 
The overall factor $\tilde \mu$ is multiplied for convenience to realize the background flat FLRW universe in the limit $\tilde \mu=0$. 
The value of $l$ is fixed as $k^{-1}=10H_{\rm b}^{-1}$ as before, and the value of $\chi_a$ is set as $\pi \tilde \mu/k$ for a given value of $\tilde \mu$. 
This profile looks complicated and differs from the original three-zone model profile but is useful for showing a specific example of the non-monotonic behavior of the PBH mass.

\begin{figure}[h]
    \centering
    \hspace{-60pt}
    \begin{minipage}[b]{0.45\linewidth}
        \centering
        \includegraphics[width=1.2\linewidth]{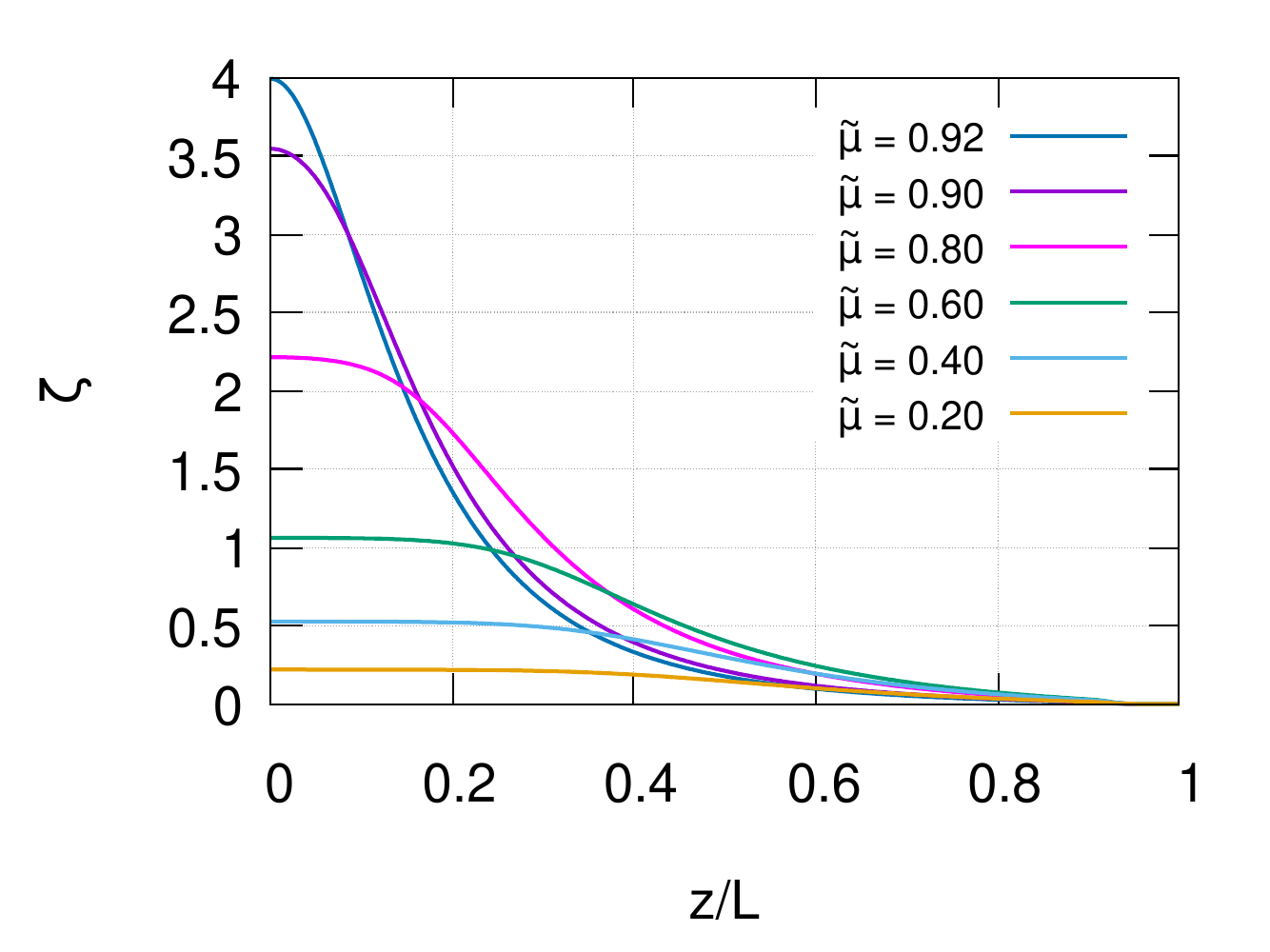}
    \end{minipage}
    \hspace{20pt}
    \begin{minipage}[b]{0.45\linewidth}
        \centering
        \includegraphics[width=1.2\linewidth]{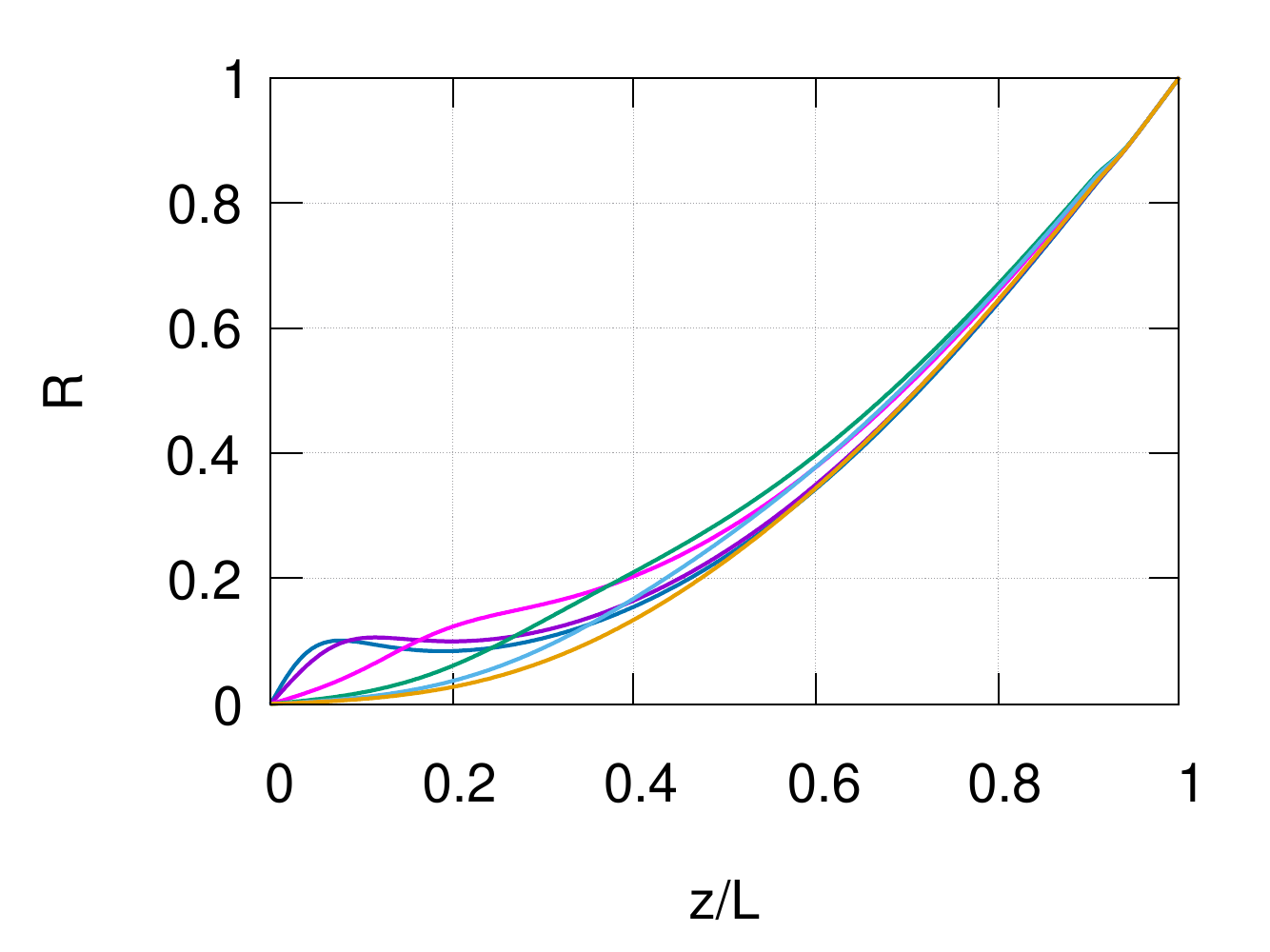}
    \end{minipage}
    \caption{
    Same as Fig.~\ref{fig:InitialGaussianProfile} but for $\zeta(r)$ given by Eq.~\eqref{eq:zetacFLRW}. The functional form~\eqref{eq:zetacFLRW} is inspired by the three-zone model but modified for numerical simulations and is different from the Gaussian-shaped profile~\eqref{eq:Gprofile}. We plot $\zeta(r)$ for the set of the amplitude parameter values $\tilde{\mu}=0.20, 0.40, 0.60, 0.80, 0.90$ and $0.92$. We can see that $R(r)$ is a monotonically increasing function for $\tilde{\mu}=0.20, 0.40, 0.60$ and $0.80$, while it increases, decreases and increases again as $r$ increases for $\tilde{\mu}=0.90$ and $0.92$. In the latter cases, $R(r)$ has a local maximum and a minimum with $\partial_{r}R=0$, where the minimum corresponds to the neck.
    }
    \label{fig:Initial3zoneProfile}
\end{figure}

\begin{figure}[h]
    \centering
    \hspace{-60pt}
    \begin{minipage}[b]{0.45\linewidth}
        \centering
        \includegraphics[width=1.2\linewidth]{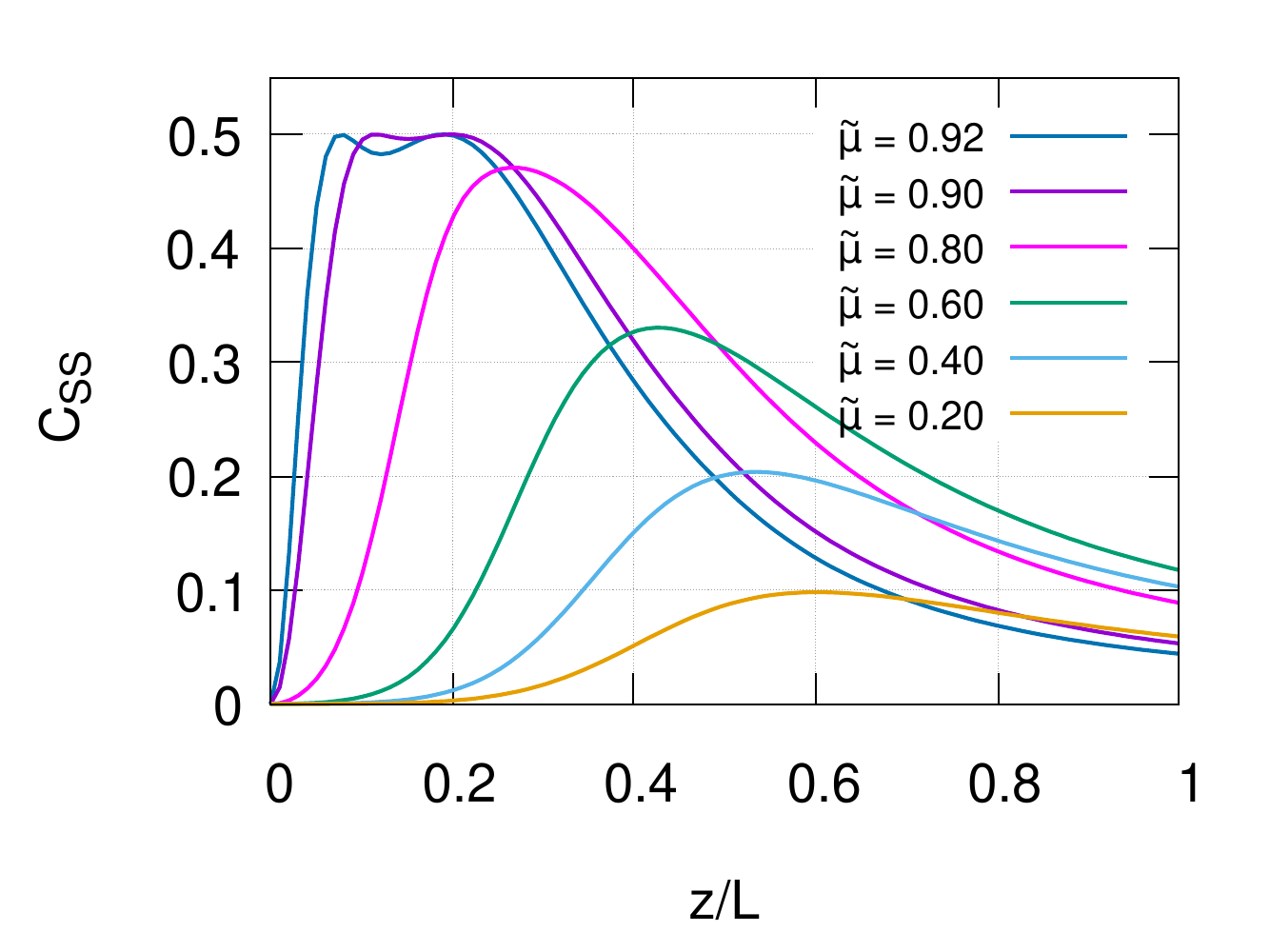}
    \end{minipage}
    \hspace{20pt}
    \begin{minipage}[b]{0.45\linewidth}
        \centering
        \includegraphics[width=1.2\linewidth]{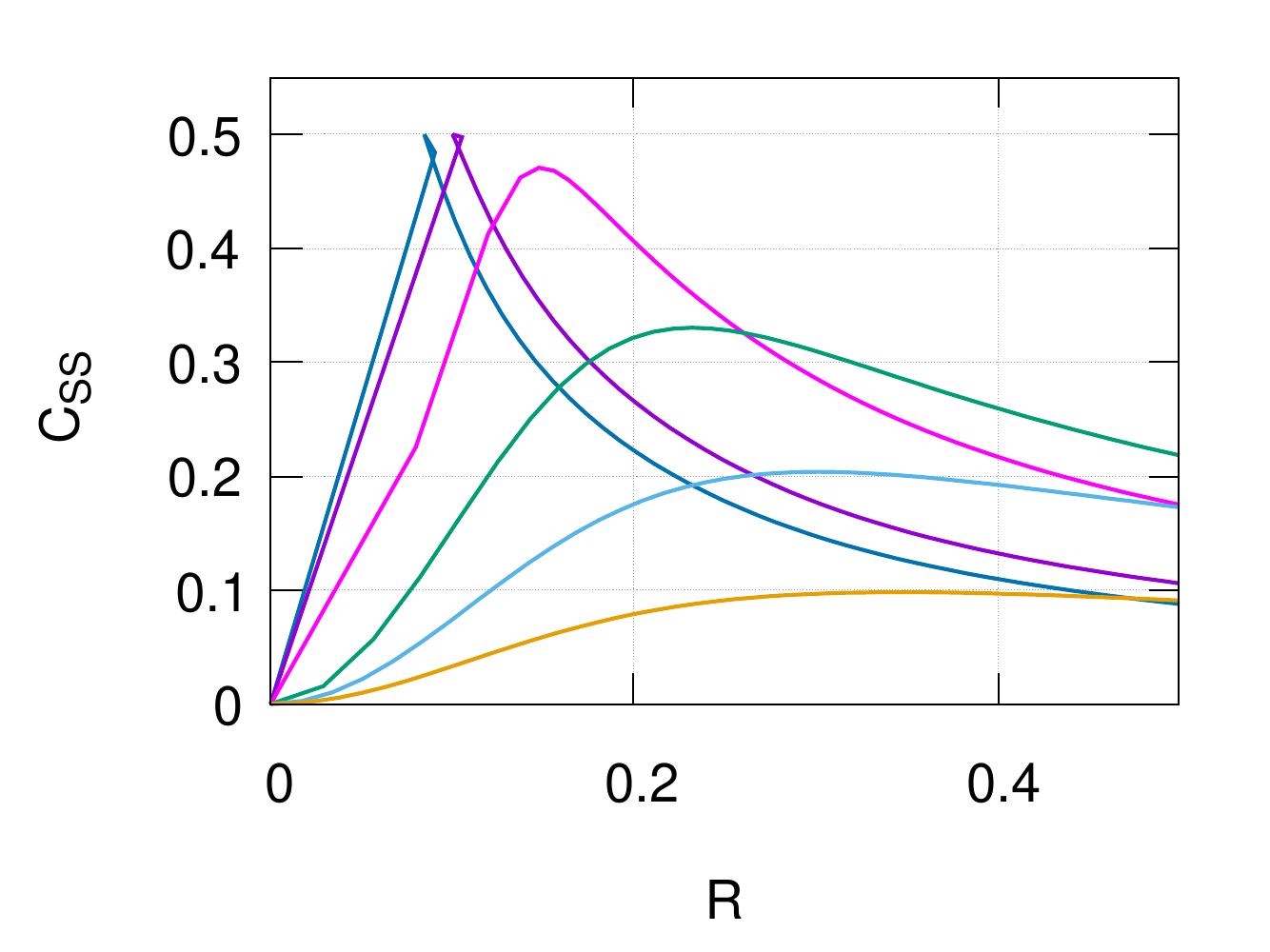}
    \end{minipage}
    \caption{
    Same as Fig.~\ref{fig:compfunc} but for $\zeta(r)$ given by Eq.~\eqref{eq:zetacFLRW}. We can see that $\mathcal{C}_\text{SS}(r)$ takes the only one maximum, which is smaller than $1/2$, for $\tilde{\mu}=0.20, 0.40, 0.60$, and $0.80$, while it takes two distinct maxima, which are equal to $1/2$, and a single minimum between the two maxima for $\tilde{\mu}=0.90$ and $0.92$. The latter cases correspond to type II fluctuations as shown in Appendix~\ref{sec:twoPeaks}. 
    We can see that $\mathcal{C}_\text{SS}(r)$ takes the only maximum or the two maxima at the smaller areal radius $R$ for the larger values of $\tilde{\mu}$. This implies that the mass of the formed black hole decreases as $\tilde{\mu}$ increases as seen in Fig.~\ref{fig:mass_amp3}. This is not the case for the Gaussian-shaped functional forms for $\zeta(r)$ shown in Fig.~\ref{fig:compfunc}, for which the mass of the formed black hole monotonically increases as $\tilde{\mu}$ increases as shown in Fig.~\ref{fig:mass_amp} at least for the parameters computed here.
    }
    \label{fig:compfunc3}
\end{figure}

\begin{figure}[h]
    \centering
    \includegraphics[width=.8\linewidth]{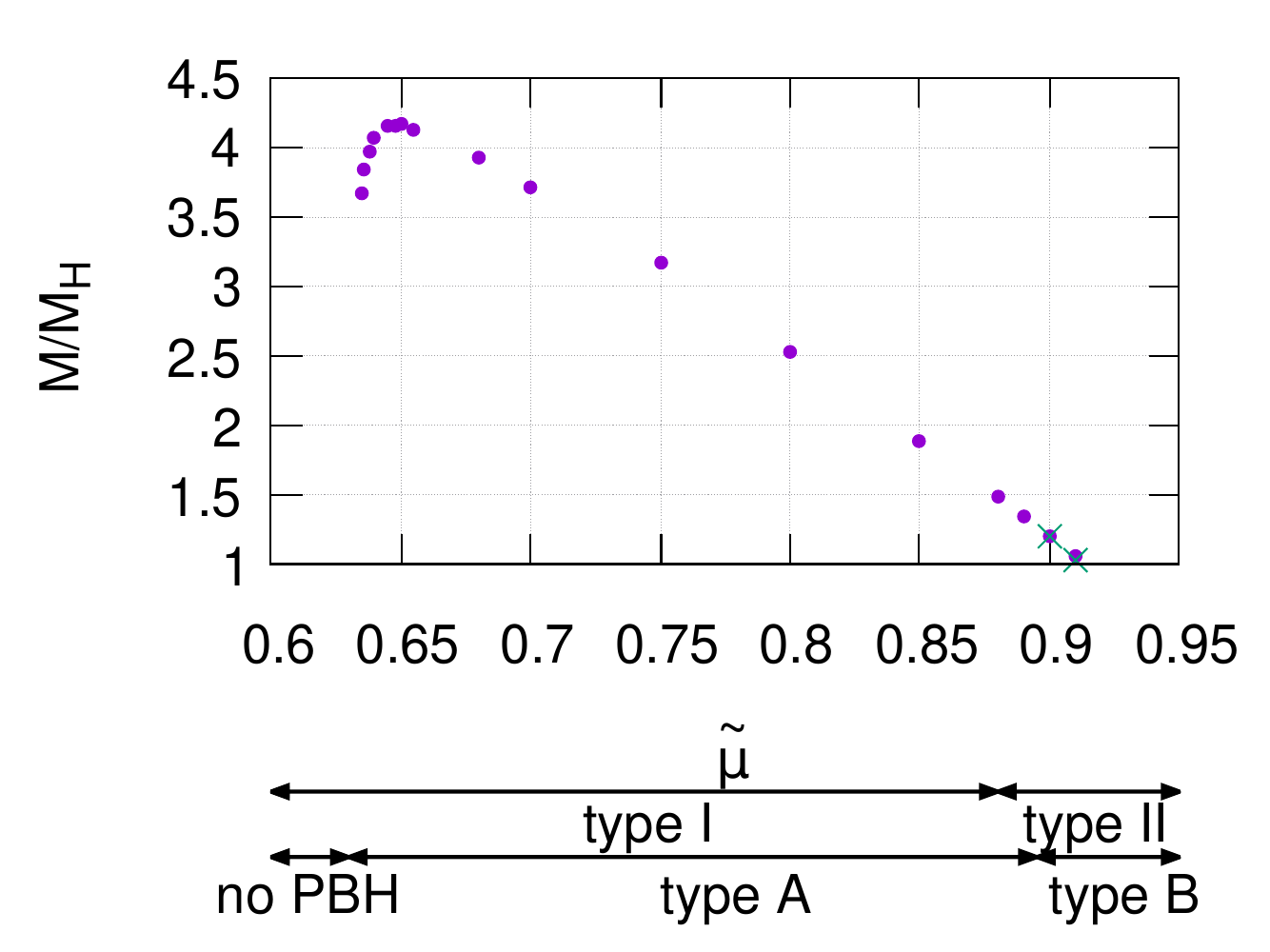}
    \caption{
        Initial PBH mass with respect to the amplitude of fluctuation for $\zeta(r)$ given by Eq.~\eqref{eq:zetacFLRW}
        We can see that the initial PBH mass decreases as the amplitude parameter $\tilde{\mu}$ is increased for $\tilde{\mu}\gtrsim 0.65$, after it increases for $\tilde{\mu}_{c}<\tilde{\mu}\lesssim 0.65$. 
    }
    \label{fig:mass_amp3}
\end{figure}

The initial profiles of $\zeta$, $R$ and the compaction function $\mathcal C_{\rm SS}$ are shown in Figs.~\ref{fig:Initial3zoneProfile} and \ref{fig:compfunc3}. 
The tail profile is different from the original three-zone model, but it is similar to the three-zone profile near the peak 
of $\zeta$.
This profile can also exhibit the type II feature with the neck structure for a large amplitude $(\tilde{\mu} \gtrsim 0.85 )$.
The behavior of the horizon configuration is similar to the case of the Gaussian profile, and we do not explicitly show it. 
However, in contrast to the Gaussian profile, the initial PBH mass formed by this profile decreases as the amplitude of fluctuation $\tilde{\mu}$ increases for $\tilde{\mu}>0.65$ as shown in Fig.~\ref{fig:mass_amp3}, even though they are classified as type I fluctuations and type A formations for $\tilde{\mu}\leq 0.88$. 
Again, this is well beyond the near-critical regime.
This result, as well as the behavior of the Gaussian profile, indicates that whether the initial PBH mass increases or decreases as a function of the amplitude parameter is not only up to its type but also strongly dependent on its profile, together with the ambiguity in quantifying the amplitude of non-linear perturbation. 
We also find the gap between the parameter region of type I/II and type A/B PBHs, as indicated in Fig.~\ref{fig:mass_amp3}. 
The decreasing behavior of the initial PBH mass with the amplitude can be easily predicted by focusing on the profile of the compaction function as a function of the areal radius. 
The initial PBH mass would be roughly estimated by the areal radius at the peak of the compaction function at the horizon entry. 
Therefore, if the areal radius at the peak of the compaction function increases and decreases with increasing the amplitude of the perturbation profile, the initial PBH mass is expected to increase and decrease as well, respectively. 
In the case of the Gaussian profile (\ref{eq:Gprofile}), one can find that the areal radius at the peak of the compaction function increases with increasing the value of the amplitude $\mu$ as seen in the right panel of Fig.~\ref{fig:compfunc}, while it decreases with increasing $\tilde \mu$ in the case of the specific profile~\eqref{eq:zetacFLRW} as seen in the right one of Fig.~\ref{fig:compfunc3}. 

\section{Conclusion}\label{sec:conclution}
We have numerically investigated the PBH formation from a type II initial fluctuation $\zeta(r)$, which is defined by the non-monotonicity of the areal radius $R(r) \propto r e^\zeta $ with respect to the coordinate radius $r$ on the spatial geometry, i.e., the existence of $\partial_r R(r)=0$ (called neck or throat structure), in the radiation-dominated situation.
As in the dust case reported in KHW, the initial fluctuation can be classified into types I and II according to the configuration of the trapped regions $R(t,r)\leq 2M(t,r)$ (or trapping horizons $R(t,r)= 2M(t,r)$). 
Specifically, if there is a bifurcating trapping horizon, the PBH is classified as type B, and otherwise, it is classified as type A.
In the dust case, KHW reported that a type I fluctuation results in a type A PBH and a type II fluctuation in a type B PBH.
However, in the case of radiation fluids, this is not true. 
We found a gap between the classifications regarding the shape of the initial spatial geometry (fluctuation type) and the structure of the trapping horizons (formation type). 
We found that a type II fluctuation may result in a type A PBH, which we will call type II-A PBH.
Therefore, we propose classifying type A/B PBHs based on the existence of the bifurcating trapping horizon independently of the type of the initial fluctuation. 

In the leading order of the long-wavelength limit, the overdense region can be roughly approximated by the closed FLRW universe, and the amplitude of the density perturbation can be translated into the value of the radial coordinate at the edge of the closed FLRW region. 
Then, the type II fluctuation indicates that the edge is located in the other hemisphere of $S^3$ beyond the great circle on the initial hypersurface. 
Therefore, in the case of type II PBHs, one may expect that the PBH mass would decrease with increasing the amplitude of initial fluctuation since the edge of the overdense region in $S^{3}$ closes in the limit of large-density perturbation, which indicates that the area of the edge is vanishing.  
However, this is not the case for the Gaussian-type initial fluctuation profile, whereas the other profile exhibits decreasing mass with increasing initial amplitude. 
Therefore, we conclude that the behavior of the initial PBH mass depends on not only the type of fluctuation but also the fluctuation profile.
Thus, a more systematic investigation would be needed to reveal the details of the profile dependence, e.g., by using the shape parameter defined as the curvature peak of the compaction function $\mathcal{C}_\text{SS}$ \cite{Escriva:2019phb,escriva_2021}.
In addition to profile dependence, dependence on the equation of state $p=w\rho$ is also an interesting issue, filling the gap between KHW and this work \cite{escriva_2021analy}.

\begin{acknowledgments}
The authors are grateful to I. Musco, K.-I. Nakao, B.J. Carr, T. Igata, and J.M.M. Senovilla for useful comments and fruitful discussion.
A.E. acknowledges support from the JSPS Postdoctoral Fellowships for Research in Japan (Graduate School of Sciences, Nagoya University).
The authors are supported in part by the JSPS KAKENHI Grant Nos. 19K03876(TH), 20H05850(CY), 20H05853(TH, CY), and 24KJ1223(DS).
KU would like to take this opportunity to thank the “THERS Make New Standards Program for the Next Generation Researchers” supported by JST SPRING, Grant Number JPMJSPS2125.
CY and TH acknowledge the hospitality at APCTP during the focus research program 
"Black Hole and Gravitational Waves: from modified theories of gravity to data analysis", where part of this work was discussed. 
\end{acknowledgments}

\appendix

\section{Conformal diagrams of the three-zone model for the dust case}\label{sec:3zonePonchi}
In the three-zone model~\cite{PhysRevD.88.084051}, the spacetime of PBH is described by the patchwork of the three different regions: the innermost overdense region is given by a spatially closed FLRW universe, the intermediate underdense region, and the outermost flat FLRW universe as the background.
We also note that the intermediate underdense region must be introduced to compensate for the innermost region's mass excess and connect the spacetime to the background spacetime without a mass gap. 
Without the intermediate underdense region, this model cannot be an exact solution, even for the dust case.

For the underdense region with a perfect fluid with $w=1/3$, obtaining any analytical and exact solutions of such dynamical inhomogeneous spacetimes with non-linear sound waves would not be tractable. 
Therefore, we stick to the three-zone model with dust. 
Furthermore, for simplicity, we focus on the model where the underdense region is described by the Schwarzschild solution, where the matching is possible without any singular hypersurface for the dust case. 
Despite this significant simplification, the model provides a powerful tool for capturing the overall dynamics of PBH and the structure of trapping horizons.
Note that one can also approach the same problem by using the Lema\^{i}tre-Tolman-Bondi exact dust solution (c.f.~\cite{PhysRevD.83.124025}).

\begin{figure}[h]
    \centering
    \hspace{-15pt}
    \begin{minipage}[b]{0.25\linewidth}
        \centering      
        \includegraphics[width=.75\linewidth,page=7]{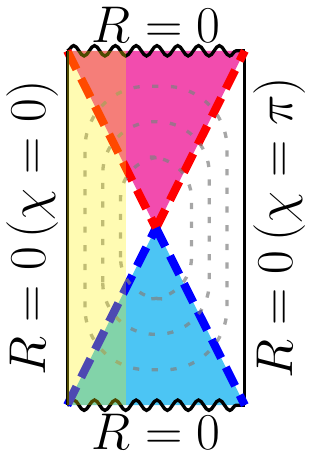}
        \subcaption{
            closed FLRW for dust
        }
    \end{minipage}
    \hspace{10pt}
    \begin{minipage}[b]{0.4\linewidth}
        \centering
        \includegraphics[width=\linewidth,page=4]{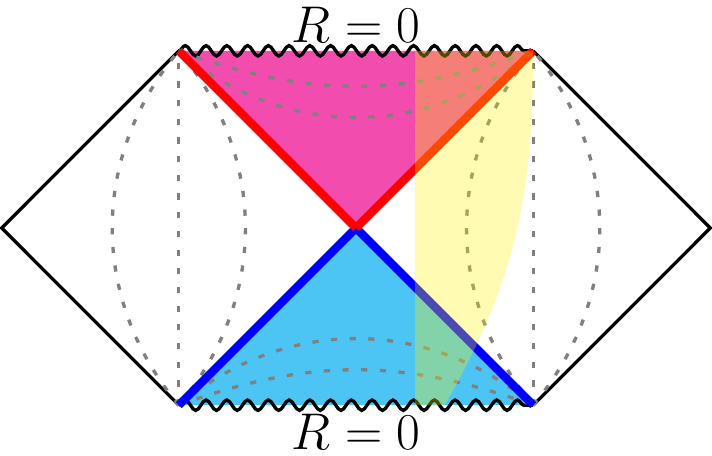}
        \subcaption{
            Schwarzschild
        }
    \end{minipage}
    \hspace{10pt}
    \begin{minipage}[b]{0.3\linewidth}
        \centering
        \includegraphics[width=.9\linewidth,page=9]{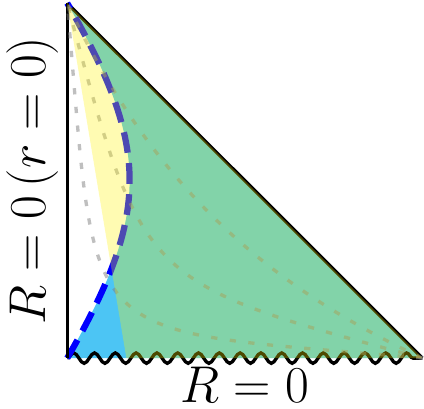}
        \subcaption{
            flat FLRW for dust
        }
    \end{minipage}
    \caption{
        Conformal diagrams to use for a patchwork for the spacetime of the type I (type A) PBH schematically in Fig. \ref{fig:PBHponchiDust} for dust case $w=0$.  
        The left, middle, and right panels show the conformal diagrams of the closed FLRW, the Schwarzschild, and the flat FLRW spacetimes, respectively.
        Dashed gray curves represent $R$-levels.
        Red (blue)-shaded regions represent future (past) trapped regions.
        Future-outer, past-outer, future-inner, and past-inner trapping horizons are denoted by red-solid, blue-solid, red-dashed, and blue-dashed lines, respectively. 
        See Table~\ref{table:table_FOTH} for the classification of trapping horizons.
    }
    \label{fig:diagrams}
\end{figure}

The conformal diagrams for the three spacetimes of which the three-zone model consists are shown in Fig.~\ref{fig:diagrams}, where future, past, and bifurcating trapping horizons are denoted by red lines, blue lines, and the intersection point between them in the Schwarzschild region, respectively.
For the dust case $w=0$, the apparent horizon and cosmological apparent horizon become future-outer and past-inner, denoted by red-solid and blue-dashed lines, respectively.
See Table~\ref{table:table_FOTH} for the classification of trapping horiozons.

\begin{figure}[h]
    \centering
    \begin{minipage}[b]{0.32\linewidth}
        \centering
        \includegraphics[height=5cm,page=10]{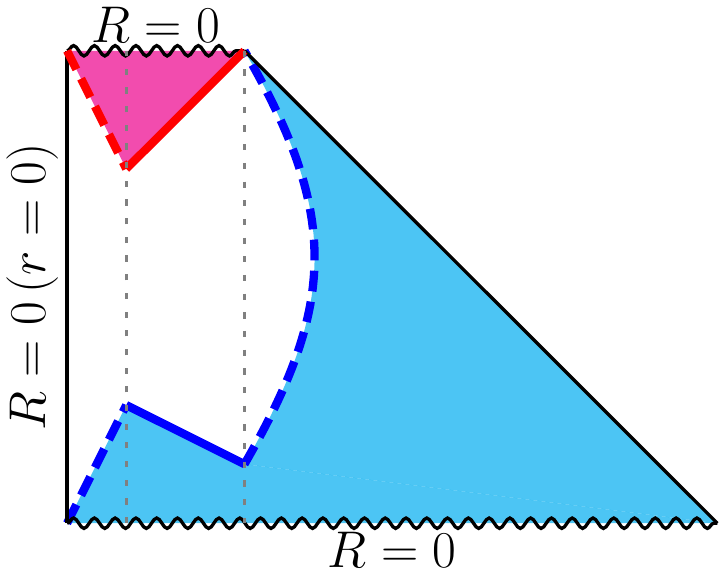}
        \subcaption{Type A PBH}
    \end{minipage}
    \hfill
    \begin{minipage}[b]{0.64\linewidth}
        \centering
        \includegraphics[height=5cm,page=11]{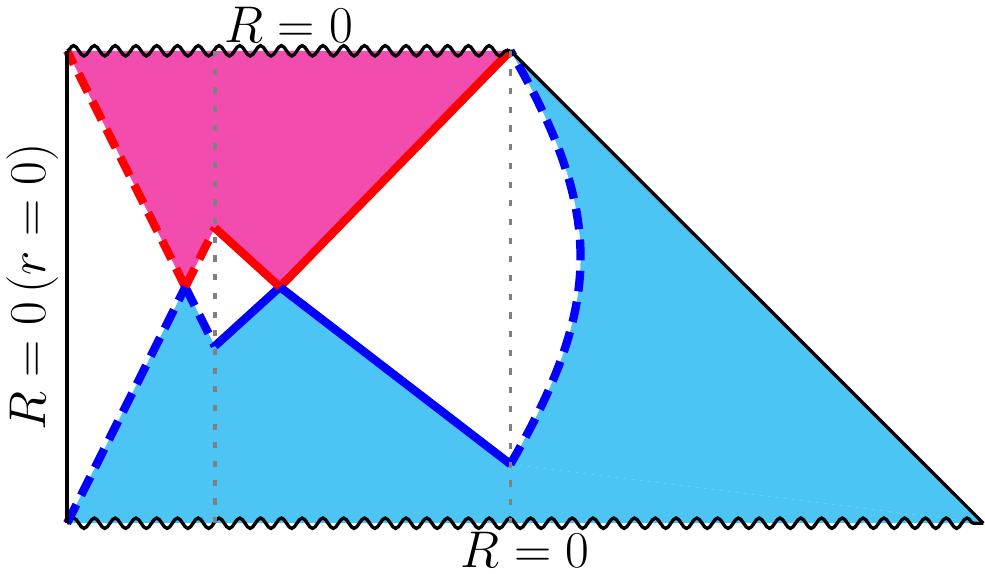}
        \subcaption{Type B PBH}
    \end{minipage}
    \caption{
        Conformal diagrams of PBH for dust case $w=0$ as a patchwork of the three diagrams and trapping horizons.
        The left panel defines a type I (type A) PBH as a typical case PBH causal structure.
        The right one is that for a type II (type B) PBH.
    }
    \label{fig:PBHponchiDust}
\end{figure}

First, let us construct the spacetime diagram of the case of the type I fluctuation. 
The extracted regions for the patchwork in the type I (type A) case are described by the yellow shaded regions in Fig.~\ref{fig:diagrams}, and the resultant spacetime diagram is shown in the left panel of Fig.~\ref{fig:PBHponchiDust}. 
For the closed FLRW region, on the boundary of the extracted region, we can find $\theta <0$ and $\theta' >0$ at the maximum expansion.  
Therefore, in the outside Schwarzschild region, we also require the same sign of $\theta$ and $\theta'$. 
Similarly, one can understand the matching of the Schwarzsdhild region and the background spatially flat FLRW spacetime.  

\begin{figure}[h]
    \centering
    \hspace{-15pt}
    \begin{minipage}[b]{0.25\linewidth}
        \centering
        \includegraphics[width=.75\linewidth,page=8]{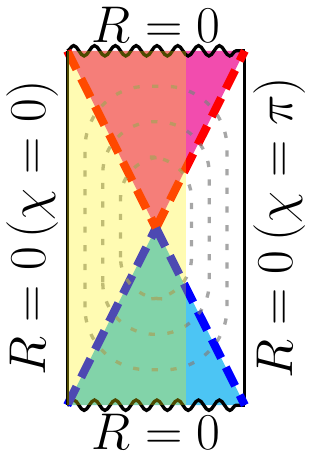}
       \subcaption{
            closed FLRW for dust
        }
    \end{minipage}
    \hspace{10pt}
    \begin{minipage}[b]{0.4\linewidth}
        \centering
        \includegraphics[width=\linewidth,page=6]{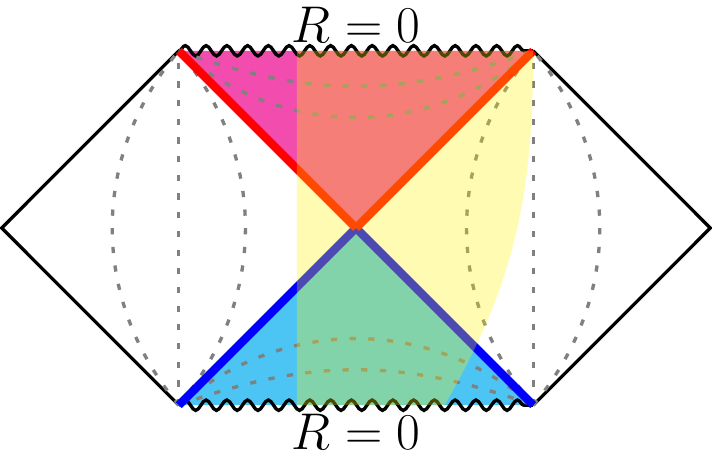}
        \subcaption{
            Schwarzschild
        }
    \end{minipage}
    \hspace{10pt}
    \begin{minipage}[b]{0.3\linewidth}
        \centering
        \includegraphics[width=.9\linewidth,page=9]{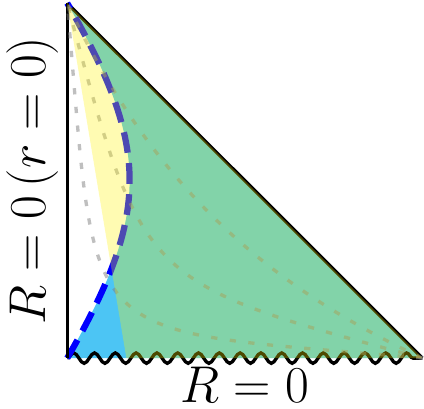}
        \subcaption{
            flat FLRW for dust
        }
    \end{minipage}
    \caption{
        Same as Fig.~\ref{fig:diagrams} but for the type II (type B) PBH. We use the yellow-shaded regions for a patchwork. 
    }
    \label{fig:diagrams2}
\end{figure}

The extracted regions are yellow-shaded for the type II fluctuation in Fig.~\ref{fig:diagrams2}. 
Since the type II fluctuation should have a maximal value of the areal radius, the boundary of the closed FLRW region must be outside the central point at which the areal radius takes the maximum value corresponding to the area of the big sphere at the maximum expansion. 
Then we find $\theta >0$ and $\theta' <0$ on the boundary, which can be matched with the Schwarzschild region only in the left part of the Schwarzschild conformal diagram. Thus, the resultant conformal diagram for type II (type B) is given by the right panel of Fig.~\ref{fig:PBHponchiDust}.

\begin{figure}[h]
    \centering
    \hspace{-20pt}
    \begin{minipage}{0.35\linewidth}
        \centering
        \includegraphics[width=.75\linewidth,page=3]{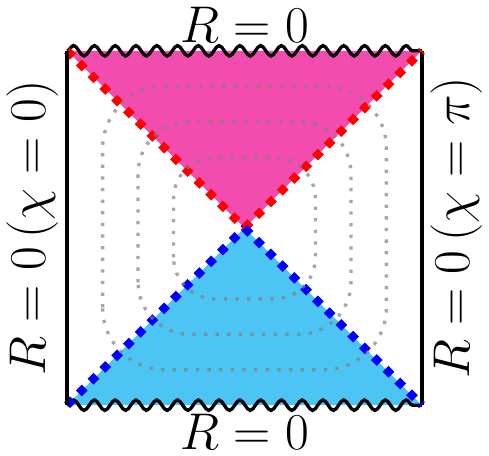}
       \subcaption{
            closed FLRW for w=1/3
        }
    \end{minipage}
    \hfill
    \begin{minipage}{0.3\linewidth}
        \centering
        \includegraphics[width=.9\linewidth,page=5]{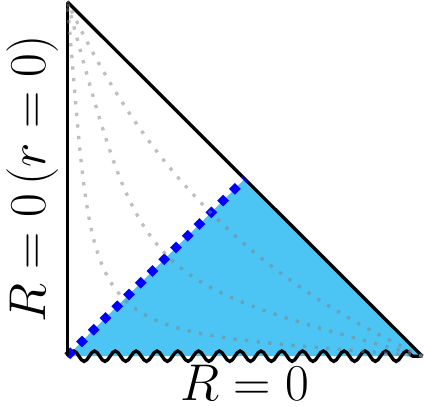}
        \subcaption{
            flat FLRW for w=1/3
        }
    \end{minipage}
    \caption{
        Conformal diagrams of the closed and flat FLRW solutions in the left and right panels, respectively, for the radiation fluid case $w=1/3$.
        In this case, all trapping horizons are null.
        Future-degenerate and past-degenerate trapping horizons are denoted by red-dotted and blue-dotted lines, respectively. 
        See Table~\ref{table:table_FOTH} for the classification of trapping horizons.
    }
    \label{fig:TOYdiagrams}
\end{figure}

For reference, Fig.~\ref{fig:TOYdiagrams} gives the conformal diagrams of the closed and flat FLRW spacetimes with $w=1/3$ in the left and right panels, respectively.

\section{Derivation of $\zeta_\text{cFLRW}$ in radial coordinate $r$}\label{sec:zeta_KHW_derivation}  
\begin{figure}[h]
    \centering
    \includegraphics[width=.8\linewidth]{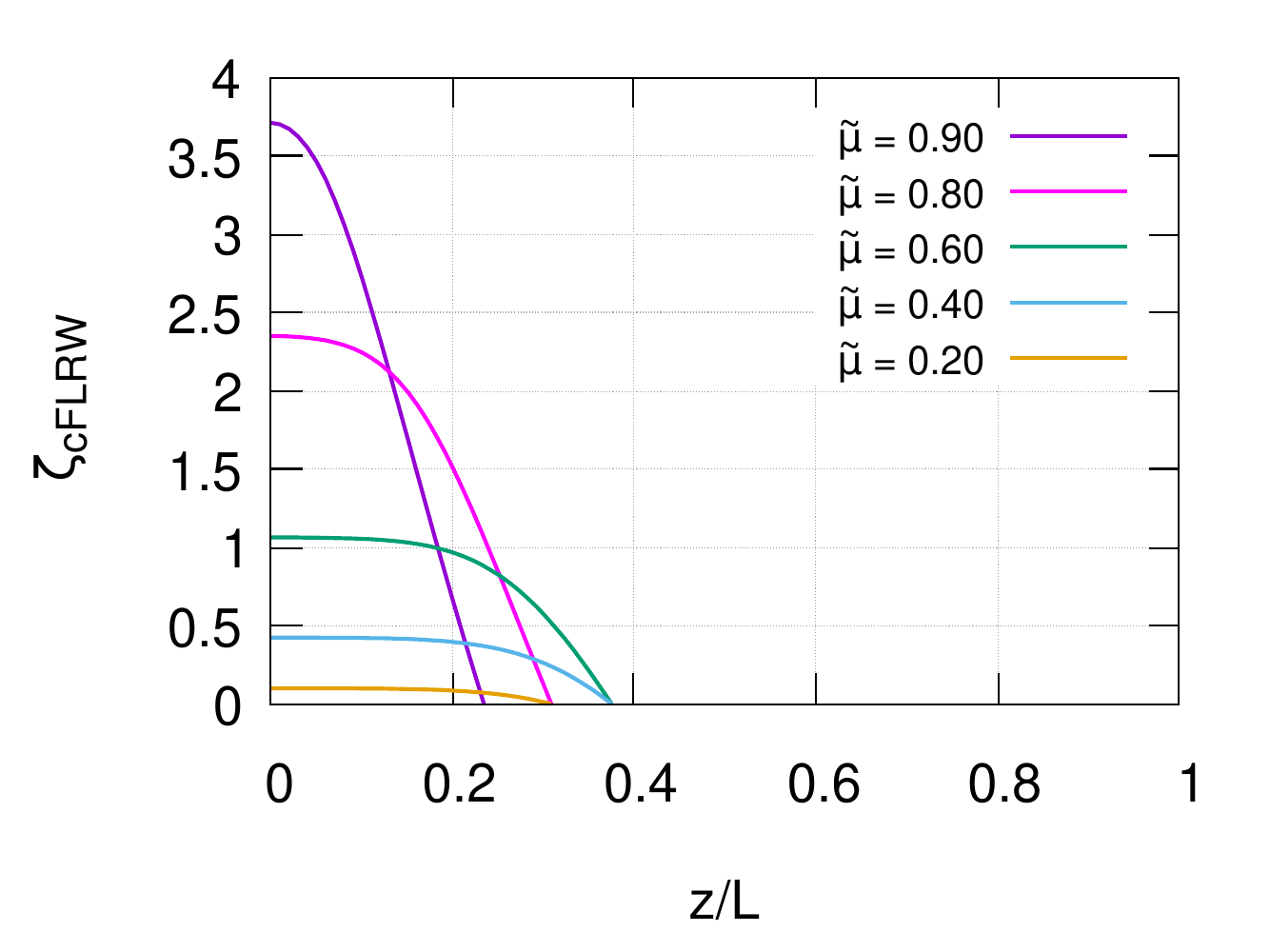}
    \caption{
        Functional form of $\zeta_\text{cFLRW}$ for overdense region $\chi\leq\chi_a$. 
        This profile is obtained by imposing a boundary condition that $\zeta=0$ at the edge of the overdense region $\chi=\chi_a$ \cite{PhysRevD.83.124025}. 
    }
    \label{fig:KHWProfile}
\end{figure}

The leading order of the long-wavelength solution in terms of spatial 3-metric is expressed as
\begin{equation}
    dl^2 =  a^2 e^{2\zeta(r)}\qty(dr^2 + r^2 d\Omega^2),
\end{equation}
where $r$ is a radial coordinate in the conformally flat geometry.
We consider this perturbed metric of an overdense region as closed FLRW geometry with a constant positive curvature $k^2$,
\begin{align}
    dl^2 &= a^2 \qty(d\chi^2 + \frac{1}{k^2}\sin^2 k\chi \, d\Omega^2),
\end{align}
where $\chi$ means the comoving radial coordinate.
By comparing the correspondence of the metric, we obtain
\begin{align}
    d\chi &= \pm dr e^\zeta,\\
    \frac{1}{k}\sin k\chi &= r e^\zeta,
\end{align}
then we get the relation,
\begin{align}
    R 
    &= \frac{a}{k}\sin k\chi = a r e^\zeta,\\
    r &= C \exp \int \frac{d(k\chi)}{\sin k\chi} = C \tan \frac{k\chi}{2},
\end{align}
where C is an integral constant.
By using these relations, $\zeta$ is obtained as
\begin{align}
    e^\zeta &=  \frac{\sin k\chi}{k r},\\
    &= \frac{2}{C k} \cos^2 \frac{k\chi}{2},\\
    &= \frac{2}{k} \frac{1/C}{1+\qty(r/C)^2}.
\end{align}
If we write $C$ as $2 C'/k$ for simplicity,
\begin{equation}
    e^\zeta = \frac{1/C'}{1+\frac{k^2}{4}\qty(r/C')^2} \quad\therefore\quad \zeta=\ln \frac{1/C'}{1+\frac{k^2}{4}\qty(r/C')^2}.
\end{equation}
Here we impose the boundary condition that $\zeta=0$ for $\chi=\chi_a$, we get
\begin{equation}
    C' = \cos^2 \frac{k \chi_a}{2}. 
\end{equation}
Then
\begin{equation}
    \zeta = 2\ln \frac{\cos k \chi_a/2}{\sqrt{\cos^4  k \chi_a/2 + \frac{1}{4}k^2 r^2}}.
\end{equation}
This functional form is shown as in Fig.~\ref{fig:KHWProfile}.


\section{
Double-peaked compaction function and averaged density perturbation in the type II fluctuation}\label{sec:twoPeaks}
\begin{figure}
    \centering
    \includegraphics{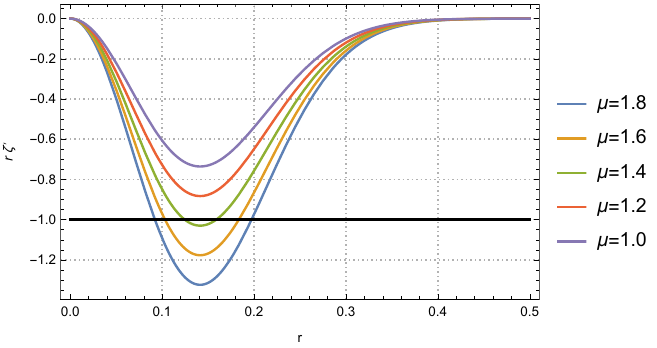}
    \includegraphics{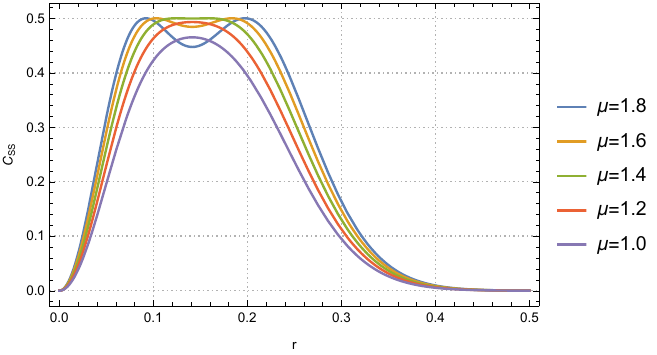}
    \caption{
    The functions $r\zeta'(r)$ and the compaction function ${\mathcal C}_{\rm SS}(r)$ are plotted in the top and bottom panels, respectively, for $\mu=1.0,1.2,1.4,1.6$ and $1.8$. 
    We can see that for $\mu=1.4, 1.6$ and $1.8$, $r\zeta'(r)$ gets smaller than $-1$, corresponding to the two peaks of the value $1/2$ of the compaction function $\mathcal{C}_{\rm SS}$ (or averaged density perturbation $\bar{\delta}$ at the horizon entry), while this is not the case for $\mu=1.0$ and $1.2$.}
    \label{fig:enter-label}
\end{figure}

For type II fluctuations, let us assume that $R'=\partial_{r}R$, where $R\coloneqq a r e^\zeta$ is the areal radius, satisfies the following condition
\begin{eqnarray}
R' 
\begin{cases}
>0 & (0<r<r_{m1}, r_{m2}<r)  \\
=0 & (r=r_{m1}, r_{m2})  \\
<0 & (r_{m1}<r<r_{m2}) 
\end{cases}
.
\end{eqnarray}
Since 
\begin{align}
R'=a e^{\zeta}(1+r\zeta'),
\end{align}
we find $r\zeta'>-1$ for $0<r<r_{m1}$, $r_{m2}<r$, $r\zeta'=-1$ for $r=r_{m1}, r_{m2}$, and $r\zeta'<-1$ for $r_{m1}<r<r_{m2}$. 
From the above assumption, we can deduce that $r\zeta'$ must have at least one minimum below $-1$ with $(r\zeta')'=0$ between $r_{m1}$ and $r_{m2}$.
For simplicity, we additionally assume within this section that $r\zeta'$ takes only one minimum at $r=r_{m3}$ between $r_{m1}$ and $r_{m2}$ and $r\zeta'$ monotonically decreases for $0<r<r_{m3}$ and monotonically increases for $r_{m3}<r$.
Since we have the compaction function \footnote{
    One can rewrite $C_{\rm SS}(r)$ as
    \begin{equation*}
        C_{\rm SS}(r) = - r \zeta'\qty(1+ \frac{1}{2} r \zeta')=\frac{3}{4}\delta_{\rm l}\qty(1-\frac{3}{8}\delta_{\rm l})
        =-\frac{9}{32}\left(\delta_{l}-\frac{4}{3}\right)^{2}+\frac{1}{2},
    \end{equation*}
    where 
    \begin{equation*}
        \delta_{\rm l} \coloneqq -\frac{4}{3} r\zeta'.
    \end{equation*}
    is the "linear-order density perturbation".
    Therefore, $C_{\rm SS}$ as a function of $\delta_{\rm l}$ takes two zeroes at $\delta_{\rm l}=0$ and $\delta_{\rm l}=8/3$ and a maximum $1/2$ at $\delta_{\rm l}=4/3$. 
    The type II fluctuation is characterized by the existence of the interval in $r$ with $\delta_{l}(r)>4/3$, where $C_{\rm SS}(r)$ as a function of $r$ takes two maxima of $1/2$ at its boundaries with $\delta_{l}(r)=4/3$.
}
\begin{equation}
   \mathcal{C}_{\rm SS}(r) = \frac{1}{2}\left[1-\qty(1+r \zeta')^2\right]
\end{equation}
and its derivative 
\begin{equation}
    \mathcal{C}_{\rm SS}'(r) = -\qty(1+r \zeta')\qty(\zeta' + r\zeta''),
\end{equation}
we find that the compaction function $\mathcal{C}_{\rm SS}(r)$ has three extrema at $r=r_{m1}$, $r_{m3}$ and $r_{m2}$ with $ \mathcal{C}_{\rm SS}'(r)=0$, two of which at $r=r_{m1}$ and $r_{m2}$ correspond to two peaks with the value $1/2$, and the rest at $r=r_{m3}$ to its minimum below $1/2$.  
This feature is typically seen for type II fluctuations and actually the case for the Gaussian-shaped profile Eq.~\eqref{eq:Gprofile} as seen in
Fig.~\ref{fig:enter-label}.

The density perturbation $\delta\coloneqq \rho/\rho_{\rm b} -1$ in a sub-leading order $\order{\epsilon^2}$ of the long-wavelength solutions for uniform Hubble (constant mean curvature) slice is written as~\cite{Harada:2015yda},
\begin{align}
    \delta \simeq -\frac{4}{3} e^{-5\zeta/2}\bar{\Delta}\qty(e^{\zeta/2}) \qty(\frac{1}{a H_{\rm b}})^2,
\end{align}
where $\bar{\Delta}$ is the Laplacian of the flat geometry.
The compaction function can be identified with the volume averaged density perturbation at the horizon entry $R=1/H_{\rm b}$,
\begin{align}
    \left.\bar{\delta}\right|_{H_{\rm b}=1/R }\simeq 2\mathcal{C}_{\rm SS} .
\end{align}
Then, the plot of $\mathcal{C}_{\rm SS}$ in Fig.~\ref{fig:enter-label} can also be understood as $\bar{\delta}$ up to the factor of $2$.
For the type II fluctuations ($\mu\gtrsim1.4$), the value of $\bar \delta$ takes a minimal value at $r=\sqrt{2}/k\simeq0.14$ with $k=10$, and the minimal value decreases with increasing $\mu$.

It is worth noting that the separate universe (pinch-off) limit will not occur for this Gaussian-shaped profile of $\zeta$, while KHW considered the limit based on the three-zone model via the closed FLRW geometry of Eq.~\eqref{eq:cFLRWprofile}.

\bibliographystyle{JHEP}
\bibliography{ref}
\end{document}